\documentclass[%
 preprint,
showpacs,preprintnumbers,
 amsmath,amssymb,
 aps,
 pra,
 longbibliography,
 lengthcheck,%
]{revtex4-1}

\DeclareFontFamily{U}{mathb}{\hyphenchar\font45}
\DeclareFontShape{U}{mathb}{m}{n}{
      <5> <6> <7> <8> <9> <10> gen * mathb
      <10.95> mathb10 <12> <14.4> <17.28> <20.74> <24.88> mathb12
      }{}
\DeclareSymbolFont{mathb}{U}{mathb}{m}{n}
\DeclareMathSymbol{\curvearrowleftright}{3}{mathb}{"F2}
\DeclareMathAlphabet{\mathpzc}{OT1}{pzc}{m}{it}

\usepackage[mathcal]{euscript}
\usepackage[greek,english]{babel}
\usepackage{cases}
\usepackage{enumerate}
\usepackage{xspace}% definition of command \xspace (add at the end of definition of any command to preserve the space after it)
\usepackage{type1cm}
\usepackage{longtable}
\usepackage{graphicx}% Include figure files
\usepackage{dcolumn}% Align table columns on decimal point
\usepackage{bm}% bold math
\usepackage{hyperref}% add hypertext capabilities
\usepackage{amsthm}
\usepackage{ifthen, paralist}
\usepackage{accents}
\usepackage[cal=boondoxo]{mathalfa} %for caligraphic lowercase symbols
\usepackage{arcs}
\usepackage{color}
\usepackage{epstopdf}
\DeclareGraphicsExtensions{.eps,.png}

\newcommand{\bra}[1]{\mathop{\langle{#1}|}\nolimits}
\newcommand{\ket}[1]{\mathop{|{#1}\rangle}\nolimits}
\newcommand{\midop}[1]{\mathop{\langle #1 \rangle}\nolimits}

\newcommand{\LR}[0]{{\mathop{\leftrightarrow}}}

\newcommand{\<}[1]{}
\newcommand{\pder}[2]{\mathop{\frac{\partial #1}{\partial #2}}}
\newcommand{\der}[2]{\mathop{\frac{d #1}{d #2}}}
\newcommand{\idx}[1]{\mbox{\scriptsize #1}}

\newcommand{\Tr}{\mathop{\rm{Tr}}}

\newcommand{\Lr}[1][]{\imglr L_{\idx{re}\ifthenelse{\equal{#1}{}}{}{,#1}}}
\newcommand{\Li}[1][]{\imglr L_{\idx{im}\ifthenelse{\equal{#1}{}}{}{,#1}}}

\newcommand{\rhoI}[2][]{#1{\mathfrak{i}}\ifthenelse{\equal{#2}{}}{}{_{#2}}}

\newcommand{\dLr}[1][]{\hat{\delta L}_{\idx{re}\ifthenelse{\equal{#1}{}}{}{,#1}}}

\newcommand{\img}[1]{\accentset{\smallfrown}{#1}}
\newcommand{\imgl}[1]{\accentset{\curvearrowright}{#1}}
\newcommand{\imgr}[1]{\accentset{\curvearrowleft}{#1}}
\newcommand{\imglr}[1]{\accentset{\curvearrowleftright}{#1}}

\newcommand{\mq}{m^{\idx{Q}}}%quaternion multiplication coefficients
\newcommand{\xLambda}{\text{\textgreek{\sampi}}}%{\Sampi}
\newcommand{\qL}{\mathcal{q}}
\newcommand{\set}[1]{\boldsymbol{#1}}
\newcommand{\genphasespace}[1][]{\Sigma_{\idx{W}}^{#1}}

\newcommand{\Lqn}{\mathcal{l}}
\newcommand{\Lqnr}[1][0]{\imglr{\mathcal{l}}_{\idx{re}\ifthenelse{\equal{#1}{}}{}{,#1}}}
\newcommand{\Lqni}[1][0]{\imglr{\mathcal{l}}_{\idx{im}\ifthenelse{\equal{#1}{}}{}{,#1}}}

\newcommand{\compound}[3][]{\imgl{a^{\ifthenelse{\equal{#1}{}}{}{#1}}_{#2}\lefteqn{\phantom{b}^{\phantom{#1}}_{#3}}b^{\ifthenelse{\equal{#1}{}}{}{#1}}}\phantom{_{#3}}}

\newcommand{\Jqn}{\mathfrak{J}}
\newcommand{\Kqn}{\mathfrak{K}}

\newcommand{\Laguerre}{\mathpzc L}
\newcommand{\farg}[1]{\hspace{-3pt}{\Big[}#1{\Big]}}%compact brackets for function argument

\let\oldgather = \gather
\let\endoldgather = \endgather
\renewenvironment{gather}[0]{\par\nobreak\noindent\oldgather}{\endoldgather}
\let\oldalign = \align
\let\endoldalign = \endalign
\renewenvironment{align}[0]{\par\nobreak\noindent\oldalign}{\endoldalign}

\newcommand{\EDITBEGIN}{\renewcommand{\linenumberfont}{\normalfont\footnotesize\sffamily\color{red}\textbf}}
\newcommand{\EDITEND}{\renewcommand{\linenumberfont}{\normalfont\tiny\sffamily}}

\renewcommand{\EDITBEGIN}{}
\renewcommand{\EDITEND}{}

\begin{document}

\title{Wigner representation of the rotational dynamics of rigid tops}

\author{Dmitry V. Zhdanov}
\email{dm.zhdanov@gmail.com}
\affiliation{Department of Chemistry, Northwestern University, Evanston, IL 60208 USA}
%\author{Denis Bondar}
%\affiliation{Princeton Univ., Dept. Chem., Princeton, NJ 08544 USA}
\author{Tamar Seideman}
\email{t-seideman@northwestern.edu}
\affiliation{Department of Chemistry, Northwestern University, Evanston, IL 60208 USA}
\pacs{
%Functional analysis, quantum mechanics,
03.65.Db,
%Semiclassical theories in quantum mechanics,
03.65.Sq,
%Rigid bodies, dynamics and kinematics of,
45.40.-f
}

\begin{abstract}

We propose a methodology to design Wigner representations in phase spaces with nontrivial topology having evolution equations with desired mathematical properties. As an illustration, two representations of molecular rotations are developed to facilitate the analysis of molecular alignment in moderately intense laser fields, reaction dynamics, scattering phenomena and dissipative processes.
\end{abstract}
%\date{\today}
\maketitle

\section{Introduction\label{@SEC:Intro}}

The dynamics of complex quantum systems on the border between classical and quantum mechanics is relevant to a variety of fields, including quantum optics and information, structural analysis, studies of matter waves and mechanisms of chemical reactions (see e.g.~\cite{BOOK-Gardiner,BOOK-Suda,BOOK-Schleich,2011-Ferrie,2013-Bonnet}). The details of these dynamics can be experimentally traced with up to attosecond resolution, owing to the advances in quantum state preparation and transient probing \cite{2012-Reid,2013-Beye,2013-Lemeshko,2013-Lepine}. However, specialized models are needed to numerically access this regime. Attractive approaches are based on semiclassical propagation of the Wigner function \cite{1932-Wigner,1984-Hillery,2010-Polkovnikov,BOOK-Zachos}, including phase integral methods \cite{2010-Dittrich} and the large family of initial value representations and their related techniques and extensions \cite{2005-Kay,2015-Davidson,2013-Bonnet}.

The idea underlying all these approaches is to find a computationally efficient way to expand any given exact generator of quantum motion in a rapidly converging series \cite{2007-Pollak}. However, the mathematical form of the exact generator of motion can be substantially altered by changing the topology of the underlying configuration space \cite{1966-Cohen,2011-Ferrie,2015-Davidson}. Such structural flexibility potentially embodies wide opportunities to equip the representation with the desired properties and behavior \cite{1999-Poirier,2009-Kiyuna,2012-Veitch}. The analysis of this resource with the specific application to rotational motion of extended bodies -- a fascinating problem with many applications \cite{1983-Blanco,1987-Huber,1999-Thoss,2007-Saha,2013-Arbelo-Gonzalez}) -- constitutes the subject of the present paper.

A variety of ways to extend the original Wigner quantization ansatz to the case of rotational dynamics were suggested and analyzed \cite{1998-Luis,2002-Ruzzi,2008-Bjork,1977-Berry,1979-Mukunda,1994-Bizarro,2010-Rigas,2011-Rigas,%
1969-Pierre,1970-Nienhuis,2013-Fischer,1999-Nasyrov,2008-Klimov}, but only a few of them are applicable to unrestricted rotations of 3-dimensional bodies. The early solutions of Refs.~\cite{1969-Pierre,1970-Nienhuis}, reduce the problem to the canonical case at the cost of extending the phase space by two artificial dimensions. A variant suggested in \cite{2013-Fischer} allows to directly extract the most useful partial distributions but involves rather complicated quantum Liouville equations. Conversely, in the Nasyrov proposal \cite{1999-Nasyrov}, the equations for free symmetric and linear tops coincide with the classical ones at the expense of complicated integro-differential form for dynamical equations and common observables in the general case. Similar drawbacks also limit the utility of schemes \cite{2008-Klimov} based on the direct extension of the Stratonovich-Weyl correspondence for spin \cite{1957-Stratonovich}.

In this paper we suggest that the roots of many of the dynamical drawbacks are hidden in the employed phase space quantization procedure. The latter usually follows closely the original Wigner reasoning \cite{1932-Wigner,1981-O'Connell}, grounded on axiomatizing certain static properties of desired quasiprobability distribution \cite{1932-Wigner,1981-O'Connell} (a notable exception is the Nasyrov's scheme \cite{1999-Nasyrov}). This approach, however, lacks the tools to explicitly control the mathematical structure and complexity of the resulting dynamical equations.

Here we show that this issue can be resolved by replacing certain traditional axioms of phase space quantization with the postulates imposed on the properties of evolution equations for Wigner function. The basics of the resulting hybrid static-dynamical phase space quantization scheme are detailed in Sec.~\ref{@SEC:fundamentals}. In the subsequent sections~\ref{@SEC:Euler} and \ref{@SEC:Complete} we apply this scheme to address the problem of developing a numerically efficient phase space quantization of rotation motions. We explore two design routes by departing from the classical Euler equations and from the Liouville equations written in terms of the components of angular momenta and quaternion parameters. Correspondingly, we arrive at  two new representations. In both cases we resolve many of the mentioned drawbacks of the existing phase space quantizations but also gain a better understanding  of quantum rotations. For example, the second representation uncovers the deep physical relation between quaternions and the raising and lowering operators of the Schwinger oscillator model and also complements the dynamical picture in the Nasyrov's quantization approach \cite{1999-Nasyrov}. These findings clarify the origins of the remarkable possibility to exactly reduce the quantum Liouvillian of the free symmetric top to the classical form. We encourage readers to check the concluding section \ref{@SEC.-concl} for a brief summary of the key features and the expected advantages of the new representations in numerical simulations.

We defer to five appendices mathematical details that we expect to interest the reader but are not necessary for conveying our message.

\section{The fundamentals of generalized Wigner representations\label{@SEC:fundamentals}}

Despite being essentially different, the quantum and classical statistical mechanics operate with the same set of objects: the set of all elementary physical events (the probability space) $\Sigma$, the algebra $\cal B$ of these events and the probability measure $\cal P$ for any measurable subset in $\cal B$ \cite{BOOK-Grishanin}. Fortunately, the Hilbert space framework is fully compatible with both classical and quantum-mechanical objects \cite{1931-Koopman,1932-Neumann}. The Wigner representation exploits this fundamental fact. It is constructed by equipping the classical phase space $\Sigma$ with such an additional scalar product $(~,~)_{\idx{W}}$ that the resulted Hilbert space $\genphasespace$ can simultaneously host both classical and quantum algebras. This change formally converts both classical and quantum quantities into operators acting in $\genphasespace$. To distinguish between them we will denote the latter by the symbol $\imgl{\phantom{1}}$, preserving the ``hat'' notation $\hat{\phantom{1}}$ for operators in the ordinary configuration Hilbert space.

Compatibility with classical mechanics requires consistency of the definitions of $(~,~)_{\idx{W}}$ and (scalar-valued) classical averaging of any physical quantity $F$ over classical canonical coordinates and momenta $q_i$ and $p_i$ ($i=1...N$):
\begin{gather}\label{intro.-(F,P)}
\midop{F}{=}(F,\img{\rho})_{\idx{W}}{=}\underset{\genphasespace
}{\int...\int} F \img{\rho}d\Omega,
\end{gather}
where $d\Omega{=}dp_1...dp_Ndq_1...dq_N$ and $\img{\rho}$ denotes the generalized probability distribution in phase space called Wigner function (or Weyl symbol of density matrix). This relation should be viewed as a classical analog of the quantum equality $\midop{\hat F}{=}\Tr[\hat F\hat \rho]$ if the classical quantity $F$ is substituted by its quantum counterpart $\imgl{F}$:
\begin{gather}\label{intro.-(F,P)-qn}
\midop{\hat{F}}{=}(\imgl{F},\img{\rho})_{\idx{W}}.
\end{gather}
Since in quantum mechanics the observables and states are treated on the same footing, it is worth requiring the following traciality relation for any two states $\hat{\rho}_1$ and $\hat{\rho}_2$:
\begin{gather}\label{intro.-(P_1,P_2)=0}
(\img\rho_1,\img\rho_2)_{\idx{W}}{=}C\Tr[\hat\rho_1\hat\rho_2],~~C{=}\rm{const}.
\end{gather}
It is also natural to impose the constraint
that the images $\imgl{F}$ of quantum observables $\hat F$ remain Hermitian in $\genphasespace$:
\begin{gather}\label{intro.-O-real}
\imgl F{=}{\imgl F}^{\dag}.
\end{gather}
Eqs.~\eqref{intro.-(P_1,P_2)=0} and \eqref{intro.-O-real} imply that
\begin{gather}\label{intro.-P-real}
\img{\rho}{=}\img{\rho}^{\dagger}{=}\img{\rho}^*,
\end{gather}
is a real-valued symmetric function of phase variables.

With this, the explicit form of images $\imgl{x_i}$ and $\imgl{p_i}$ of the quantum coordinate and momentum operators (termed Bopp operators \cite{1961-Bopp,1984-Hillery}) is uniquely defined by 1) the fundamental property of Galilean invariance of non-relativistic phase space $\genphasespace$ (which requires $\imgl{x_i}$ and $\imgl{p_i}$ to be linear in both $p_i$, $q_i$ and $\pder{}{p_i}$, $\pder{}{q_i}$); 2) the canonical commutation relation $[\imgl x_i,\imgl p_i]{=}i\hbar$ and 3) the requirement of proper classical limit $\imgl{p_i}|_{\hbar\to0}{=}p_i$,  $\imgl{q_i}|_{\hbar\to0}{=}q_i$:
\begin{gather}\label{intro.-left_operators_def}
\imgl{x_i}{=}x_i{+}\frac{i\hbar}{2}\pder{}{p_i};~~~\imgl{p_i}{=}p_i{-}\frac{i\hbar}{2}\pder{}{x_i}.
\end{gather}
Note that these operators when applied to $\img \rho$ produce images of the left multiplications $\hat p_i\hat\rho$ and $\hat q_i\hat\rho$. It is convenient to introduce the operators ${\imgr{p_i}}$ and ${\imgr{q_i}}$ whose effect on $\img \rho$ is associated with the right multiplications. The associativity relations of form $\forall \hat\rho:\hat q_i(\hat\rho\hat q_j){=}(\hat q_i\hat\rho)\hat q_j$ and the equality $[\hat q_i, \hat p_i]\hat\rho{=}{-}(\hat\rho^{\dag}[\hat q_i, \hat p_i])^{\dag}$ imply that the right operators should satisfy the commutation relations:
\begin{gather}\label{intro.-right_operators_commutators}
[{\imgr{p_i}},\imgl{p_j}]{=}[{\imgr{p_i}},\imgl{q_j}]{=}[{\imgr{q_i}},\imgl{p_j}]{=}[{\imgr{q_i}},\imgl{q_j}]{=}0\\
[{\imgr{q_i}},{\imgr{p_j}}]{=}{-}i\delta_{i,j}\hbar.\notag
\end{gather}
Combining Eq. \eqref{intro.-right_operators_commutators} with the requirements of Galilean invariance and proper classical limit one can conclude that:
\begin{gather}\label{intro.-right_operators_def}
\imgr{p_i}{=}x_i{-}\frac{i\hbar}{2}\pder{}{p_i}{=}\imgl{p_i}^{*};~~~\imgr{q_i}{=}p_i{+}\frac{i\hbar}{2}\pder{}{x_i}{=}\imgl{q_i}^{*}.
\end{gather}

The equality $\Tr[\hat q_i^n\hat\rho] = \frac1{2^n} \sum_{m=0}^nC_n^m\Tr[\hat q_i^m\hat\rho\hat q_i^{n-m}]$, where $C_n^m$ are binomial coefficients, and the similar expression for $\hat p_i$ lead to the conclusion that
\begin{gather}
(\imgl{q_i}^n,\img{\rho})_{\idx{W}}{=}\frac1{2^n}(({\imgl{q_i}+\imgr{q_i}})^n,\img{\rho})_{\idx{W}}{=}(q_i^n,\img{\rho})_{\idx{W}};\notag\\
(\imgl{p_i}^n,\img{\rho})_{\idx{W}}{=}(p_i^n,\img{\rho})_{\idx{W}}\label{intro.-<q^n>,<p^n>}.
\end{gather}
In particular, Eqs.~\eqref{intro.-<q^n>,<p^n>} mean that the partial integration on the right-hand side of \eqref{intro.-(F,P)} over coordinates (momenta) with $F{=}1$ returning the correct marginal probability distributions for values of momenta (coordinates).

Equations~\eqref{intro.-left_operators_def} and \eqref{intro.-right_operators_def} completely specify the quantum algebra and establish one-to-one correspondence between an arbitrary quantum operator $\hat F{=}F(\hat{p},\hat{q})$, its Wigner image $\imgl{F}{=}F(\imgl{p},\imgl{q}))$ and the Weyl symbol $F_{\idx{W}}(p,q){=}F(\imgl{p},\imgl{q})1$ (see \cite{1984-Hillery,2010-Polkovnikov}) for details), as well as define the image of the master equation $\pder{}{t}\hat\rho{=}\frac{-i}{\hbar}[\hat H,\hat\rho]$ with Hamiltonian $\hat H{=}H(\hat{p},\hat{q})$:
\begin{gather}\label{intro.-Liouville_equation}
\pder{}{t}\img{\rho}{=}\imgl{\cal L}\img{\rho},
\end{gather}
where the quantum Liouvillian $\imgl{\cal L}$ is given by the real operator
\begin{gather}\label{intro.-Liouvillian}
\imgl{\cal L}{=}\frac{{-}i}{\hbar}(H(\imgl{p},\imgl{q}){-}H(\imgr{p},\imgr{q})).
\end{gather}

However, in the general case of non-canonic phase spaces,  Eqs.~\eqref{intro.-(F,P)-qn}-\eqref{intro.-Liouville_equation} are not self-consistent, and hence some of them must be relaxed, e.g.:
\begin{enumerate}[(I)]
\item \label{intro.-def_Wigner_stat}one can impose the desired ``static'' properties of the quasiprobablilty distribution like \eqref{intro.-(F,P)-qn}, \eqref{intro.-(P_1,P_2)=0}, \eqref{intro.-P-real}, \eqref{intro.-<q^n>,<p^n>} and then deduce from them the expressions for Weyl symbols, Moyal products and evolution equations, alternatively,
\item \label{intro.-def_Wigner_dyn}one can depart from the desired algebraic and dynamic properties of images of quantum operators and$\backslash$or generators of motion (e.g. Eqs.~\eqref{intro.-O-real},\eqref{intro.-left_operators_def},\eqref{intro.-right_operators_commutators}).
\end{enumerate}

Algorithm (\ref{intro.-def_Wigner_stat}) is rigorously axiomatized \cite{1981-O'Connell}; its abstracted generalization in group-theoretical terms (termed Stratonovich-Weyl correspondence \cite{1957-Stratonovich}) can be applied to arbitrary phase spaces with complex symmetries (see e.g. \cite{1989-Varilly,2013-Li}).
One practical and formally justified \cite{2012-Bondar,2013-Bondar} axiomatic basis for algorithm (\ref{intro.-def_Wigner_dyn}) postulates the equations of motion for averages of certain physical quantities \cite{1999-Nasyrov,1987-Dodonov}. Another possible starting point is Feynman's path integral representations of the time evolution \cite{2010-Polkovnikov}. One of the fundamental origins of this diversity of possible definitions is the wide freedom in choosing either the Weyl symbols of density matrices $\imgl\rho$ or quantum observables $\imgl F$ to be a main ``carriers of nonclassicality'' (see \cite{2011-Ferrie,2009-Ferrie} for details%theory of frames
).

\EDITBEGIN

The quantization method presented in this paper uses this diversity to construct Wigner quantizers tailored to specific dynamical problems. First, we identify the desired dynamical characteristics of the quantizer that would enhance its applied value. In our examples we consider goals such as %the preferable parameterization of the phase space,
computational simplicity of the quantum equations of motion, preferable forms of certain Bopp operators etc. At the next step we introduce these preferences into the standard set of postulates of Wigner quantization. In doing this, we have to relax some of these ``canonic'' postulates (considered as ``the least important ones'' in the context of the anticipated applications) in order to obtain a consistent axiomatic basis. The modified postulates no longer uniquely specify the representation but are accompanied by an additional ``loose'' dynamical criteria (such as ``simplicity'' of certain operators etc.) defined in physical rather than mathematical terms. This makes the construction algorithm (\ref{intro.-def_Wigner_dyn}) more suitable: one starts with deducing the forms of dynamical Bopp operators which best account for these additional criteria and then completes the definition of the Wigner function and its ``static'' properties accordingly.

Case studies illustrating this general scheme are presented in the following two sections.

\EDITEND

%pragmatic anatomize physical
%For the dynamical problem in hand.
%desired mathematical dynamical properties of the final result,
%In this paper we show that this feature allows to balance the complexities of evolution equations and the associated Wigner functions.

\section{Quantization of the Euler equations\label{@SEC:Euler}}

Formulated in 1765, Euler's celebrated equations,
\begin{gather}\label{euler.-Euler_equations}
\der{}{t}L_i{=}\sum_{j,k{=}1}^{3}\epsilon_{i,j,k}\left(\frac{1}{I_k}{-}\frac{1}{I_j}\right)\frac{L_j L_k}{2},~~~(i{=}1,2,3),
\end{gather}
where $\epsilon_{i,j,k}$ is the Levi-Civita symbol, $I_k$ are moments of inertia about the principal axes $\vec e_k$ of the rigid body and $L_k$ are the projections of the angular momentum on $\vec e_k$, describe the free dynamics of rigid bodies in the moving frame $S$ in terms of the phase space $\genphasespace[L]{=}\{L_1,L_2,L_3\}$.

We will require the quantum generalization of Eqs.~\eqref{euler.-Euler_equations} to obey:
\renewcommand{\labelenumi}{(E:\arabic{enumi})}
\noindent\begin{enumerate}\itemsep1pt \parskip0pt \parsep0pt
\item \label{euler.-postulate_P-real} the  condition \eqref{intro.-P-real} of reality of the quasiprobability distributions $\img{\rho}{=}\img{\rho}(L_1,L_2,L_3)$: any real-valued Weyl symbol $\img{\rho}'$ should correspond to unique Hermitian (but not necessarily positive) matrix $\hat\rho'$;
\item \label{euler.-postulate_(P_1,P_2)=0}the traciality relation \eqref{intro.-(P_1,P_2)=0} in $\genphasespace[L]$ where $(\img\odot_1,\img\odot_2)_{\idx{W}}\overset{\idx{def}}{=}\int_{\genphasespace[L]}\img\odot_1^*\img\odot_2dL_1dL_2dL_3$ and $C{=}1$;
\item \label{euler.-postulate_cl_limit} the consistency of the classical limit for $\imgl L_i$ and evolution equation \eqref{intro.-Liouville_equation} with the Euler equations \eqref{euler.-Euler_equations}:
\begin{gather}\label{euler.-classical_limits}
\imgl L_i|_{h{\to}0}{=}L_i;~~\imgl{\cal L}|_{\hbar{\to}0}{=}{\cal L}{=}\sum_{i,j,k{{=}1}}^{3}\epsilon_{i,j,k}(\frac{L_j L_k}{I_k})\pder{}{L_i}.
\end{gather}
\end{enumerate}

\EDITBEGIN

The postulates (E:\ref{euler.-postulate_P-real}) and (E:\ref{euler.-postulate_(P_1,P_2)=0}) match the conventional Wigner reasoning \cite{1981-O'Connell}. They also guarantee that the Hermiticity condition \eqref{intro.-O-real} holds (i.e.~$\imgl F^*|_{\pder{}{L_j}{\to}{-}\pder{}{L_j}}{=}\imgl F$ for the image $\imgl F$ of any observable). Hence, these postulates preserve the essential subset of the properties of conventional Wigner function. 
However, in defining (E:\ref{euler.-postulate_cl_limit}) we apply the algorithm \ref{intro.-def_Wigner_dyn} and substitute the static condition \eqref{intro.-<q^n>,<p^n>} on the marginal distributions with a dynamical restriction. The postulates (E:\ref{euler.-postulate_P-real}-\ref{euler.-postulate_cl_limit}) do not uniquely determine the Wigner representation. This allows to introduce the condition of mathematical simplicity of the quantum Liouvillian $\imgl{\cal L}$ as the additional ``loose'' dynamical figure of merit. 

It is worth stressing that the postulates (E:\ref{euler.-postulate_P-real}-\ref{euler.-postulate_cl_limit}) do not allow to immediately identify (i.e{.} in a way similar to the standard quantization scheme \cite{1981-O'Connell}) the explicit form of the isomorphism between the density matrix formulation and the Wigner function formulation. Instead, we have to start by identifying the structure of the key dynamical Bopp operators $\imgl L_k$ and $\imgl{\cal L}$ consistent with the given postulates and then go back and complete the identification of the function $\img\rho$ that enters eq.~\eqref{intro.-Liouville_equation}.

\EDITEND

The images of the associativity and commutation relations $\forall \rho:\hat L_k(\rho\hat L_l){=}(\hat L_k\rho)\hat L_l$; $[\hat L_l,\hat L_k]{=}i\hbar\sum_{m}{\epsilon_{k,l,m}}\hat L_m$ satisfying (E:\ref{euler.-postulate_P-real}-\ref{euler.-postulate_cl_limit}) read as:
\begin{gather}\label{euler.-[L_i,l_j]}
\forall k,l:[\imgl L_k,\imgr L_l]{=}0;~[\imgl L_k,\imgl L_l]{=}{-}i\hbar\sum_{m{=}1}^3\epsilon_{k,l,m}\imgl L_m.
\end{gather}

The postulate (E:\ref{euler.-postulate_P-real}) uniquely defines the form of any right operator $\imgr F$. Indeed, the images of the
 expressions $i[\hat F,\hat\rho']$, $[\hat F,\hat\rho']_+$ must be real for any Hermitian  $\hat\rho'$. Thus, the operators $i(\imgl F{-}\imgr F)$ and $(\imgl F{+}\imgr F)$ have to be real, i.e.:
\begin{gather}\label{euler.-right_operators_def}
\imgr F{=}\imgl F^*~~~\mbox{(cf. with \eqref{intro.-left_operators_def} and \eqref{intro.-right_operators_def})}
\end{gather}
Relation~\eqref{euler.-right_operators_def} allows to define: $\imgl L_k{=}\Lr[k]{+}i\Li[k]$, $\imgr L_k=\Lr[k]{-}i\Li[k]$ and rewrite Eqs.~\eqref{euler.-[L_i,l_j]} in the equivalent form:\begin{subequations}\label{euler.-[L_re,L_im]}
\begin{gather}
\forall k,l:[\Lr[k],\Li[k]]{=}0;\label{euler.-[L_re,L_im]=0}\\
[\Li[l],\Li[k]]{=}
[\Lr[k],\Lr[l]]{=}\frac{\hbar}{2}\sum_{m{=}1}^{3}\epsilon_{k,l,m}\Li[m];\label{euler.-[L_re,L_re]=[L_im,L_im]=L_re}\\
[\Lr[k],\Li[l]]{=}\frac{\hbar}{2}\sum_{m{=}1}^{3}\epsilon_{k,l,m}\Lr[m]\label{euler.-[L_re,L_im]=L_re}.
\end{gather}\end{subequations}

Using Eq.~\eqref{intro.-Liouvillian} and the free rigid top Hamiltonian,
\begin{gather}
\hat H{=}\sum_{k{=}1}^{3}\frac{\hat L_{k}^2}{2I_k},
\end{gather}
 one obtains,
\begin{gather}\label{euler.-Liouvillian_qn_re_im}
\imgl{\cal L}{=}2\sum_{k{=}1}^{3}\frac{\Lr[k] \Li[k]}{\hbar I_k}.
\end{gather}
Applying (E:\ref{euler.-postulate_cl_limit}) to \eqref{euler.-Liouvillian_qn_re_im} leads to the expressions for $\Li[k]$:

\EDITBEGIN

\begin{gather}\label{euler.-L_im}
\Li[k]{=}\frac 12 \hbar\sum_{i,j{=}1}^{3}\epsilon _{i,j,k}L_i\pder{}{L_j}
\end{gather}
up to terms of order $\hbar^2$ that should be chosen equal to zero in order to best satisfy our requirement of simplicity of $\imgl{\cal L}$. The system of differential equations~\eqref{euler.-L_im} and \eqref{euler.-[L_re,L_im]} can be solved for $\Lr[k]$:

\EDITEND

\begin{gather}
\Lr[k]{=}L_k{+}\frac{\hbar^2}{16}
\left(\vphantom{\int}
{-}2\sum_{i=1}^3L_i\pder{}{L_i}\pder{}{L_k}{+}\notag\right.\\
\left.L_k\sum_{i=1}^3\pder{^2}{L_i^2}{+} c_1\pder{}{L_k}{+}\xi^2\frac{L_k}{L^2}
\vphantom{\int}\right)\label{euler.-L_re},
\end{gather}
where we denoted $L{=}\sqrt{L_1^2{+}L_2^2{+}L_3^2}$. The angular momentum components defined by \eqref{euler.-L_im}, \eqref{euler.-L_re} satisfy the relations \eqref{euler.-[L_i,l_j]} and \eqref{euler.-right_operators_def} for any real values of $c_1$ and $\xi$
but only the choice $c_1{=}{-}3$ is consistent with the Hermiticity condition \eqref{intro.-O-real}. The value of $\xi$ can be selected to simplify the expression for averages originating from (E:\ref{euler.-postulate_(P_1,P_2)=0}):
\begin{gather}
\midop{\hat F}{=}(\rhoI[\img]{},\imgl F\img{\rho})_{\idx{W}},~~~\mbox{(cf. Eq.~\eqref{intro.-(F,P)-qn})}
\end{gather}
\EDITBEGIN
where $\rhoI[\img]{}$ is the Weyl symbol of the reduced identity matrix: $\rhoI[\hat]{}{=}\sum_l\rhoI[\hat]{l}$. Here $\rhoI[\hat]{l}{=}\sum_{m,k{=}{-}l}^l\hat\rho_{l,k;l,k}$ where $\hat\rho_{l,k_1;l,k_2}{=}\frac{1}{2l{+}1}\Tr_m[\ket{l,m,k_1}\bra{l,m,k_2}]$ are reduced projectors that satisfy the standard relations: $\hat L^2\hat\rho_{l,k_1;l,k_2}{=}\hbar\,l(l{+}1)\hat\rho_{l,k_1;l,k_2}$, $\hat L_3\hat\rho_{l,k_1;l,k_2}{=}\hbar\,k_1\hat\rho_{l,k_1;l,k_2}$, $\Tr[\hat\rho_{l,k_1;l,k_2}\hat\rho_{l,k_3;l,k_4}]{=}\delta_{k_1,k_4}\delta_{k_2,k_3}$. The trace $\Tr{}_m[\hat\odot]$ is taken only over the quantum number $m$ of the projection of the angular momentum on an arbitrary laboratory-fixed axis. The results of Appendix~\ref{@SEC.-app1} indicate that the following isomorphism holds between the Weyl symbols $\img\rho$ and the reduced density matrices $\hat\rho_{\idx{red}}{=}\Tr{}_m[\hat\rho]$, such that $\Tr[\hat\rho_{\idx{red}}\hat\rho_{l_1,k_1;l_2,k_2}]{=}0$ for any $l_1{\ne}l_2$ (the origin of the latter limitation will be clarified below):
\begin{subequations}\label{euler.-rho-isomorphism}
\begin{gather}
\img\rho{=}\sum_{l,k_1,k_2}Tr[\hat\rho_{l,k_2;l,k_1}\hat\rho_{\idx{red}}]\img\rho_{l,k_1;l,k_2};\\
\hat\rho_{\idx{red}}{=}\sum_{l,k_1,k_2}(\img\rho_{l,k_1;l,k_2},\img{\rho})_{\idx{W}}\hat\rho_{l,k_1;l,k_2}\label{euler.-rho-isomorphism_rev}.
\end{gather}
\end{subequations}
The general expressions for basis functions $\img\rho_{l,k_1;l,k_2}$ valid for arbitrary value of $\xi$ are given in Appendix~\ref{@SEC.-app1}. They take the most convenient non-singular form when $\xi{=}\frac12$. For this case eqs.~\eqref{app1.-|l,l><l,l|} and \eqref{app1.-i-solution;xi=1/2} give $\rhoI[\img]{}|_{\xi{=}\frac12}{=}\frac{1}{\hbar\sqrt{\pi L}}$ and:
\begin{gather}\label{euler.-image_of_|l,l><l,l|-projector}
\img\rho_{l,l;l,l}\left.\vphantom{{}_1^1}\right|_{\xi{=}\frac12}{=}\frac{4}{\hbar\sqrt{\pi L}} ({-}1)^{2 l}e^{{-}\frac{4 L}{\hbar }} \Laguerre_{2 l}\left[\frac{4(L{+}L_3)}{\hbar}\right];\\
\img\rho_{l,k_1;l,k_2}{=}c(\imgl L_1{+}i\imgl L_2)^{l{-}k_1}\hspace{-0.05cm}(\imgr L_1{-}i\imgr L_2)^{l{-}k_2}\hspace{-0.05cm}\img\rho_{l,l;l,l}\label{euler.-image_of_|l,k_1><l,k_2|-projector},
\end{gather}
where $c{=}\frac{\hbar^{k_1{+}k_2{-}2l}}{(2l)!}\sqrt{\frac{(l+k_1)!(l+k_2)!}{(l-k_1)!(l-k_2)!}}$ and the notation $\Laguerre_{2 l}$ stands for Laguerre polynomials.

\EDITEND
It is useful to highlight several peculiarities of the obtained representation.

1. In the angular momentum case, Eqs.~\eqref{intro.-(F,P)-qn} and \eqref{intro.-(P_1,P_2)=0} can not be simultaneously satisfied because of nonuniform density of quantum states in $\genphasespace[L]$: $\rhoI[\img]{}{\ne}1$. However, one may set $c_1{=}{-}4$, $\xi{=}0$ to satisfy \eqref{intro.-(F,P)-qn} instead of \eqref{intro.-(P_1,P_2)=0} which is equivalent to the non-unitary transformation,
\begin{gather}\label{euler.-sol:c_1=-4,xi=0}
\img \rho'{=}(\hbar^2\pi L)^\eta\img \rho|_{\xi{=}\frac12,c_1{=}{-}3};~~\imgl F'{=}{L}^{\eta}~\imgl F|_{\xi{=}\frac12,c_1{=}{-}3}~{L}^{{-}\eta}
\end{gather}
with $\eta{=}{-}\frac12$.
%[???!!!-novel result]

\EDITBEGIN

2. The conceptual drawback of the Wigner representations corresponding to the choices $c_1{=}{-}3$, $\xi{=}\frac12$ and $c_1{=}{-}4$, $\xi{=}0$ is that the associated Weyl symbols $\rhoI[\img]{}{=}\frac{1}{\hbar\sqrt{\pi L}}$ and $\rhoI[\img]{}'{=}\frac{1}{\hbar^2\pi L}$ are not equivalent to the Bopp operator $\rhoI[\imgl]{}{=}1$ of the identity matrix. This drawback can be eliminated by the choice $c_1{=}{-}2$, $\xi{=}0$, which corresponds to $\eta{=}\frac12$ in \eqref{euler.-sol:c_1=-4,xi=0}. The resulting formalism will be referred as $\genphasespace[L{\star}]$-re\-presentation. It is straightforward to check that $\rhoI[\img]{}^{\star}{=}\rhoI[\imgl]{}^{\star}{=}1$, so that the associated Bopp operators $\imgl O^{\star}$ and Weyl symbols  $\img O^{\star}$, $\img\rho^{\star}_{\idx{red}}$ of any reduced density matrix $\hat\rho_{\idx{red}}$ and operator $\hat O$ obey the following simple correspondence rules:\begin{subequations}\label{euler.-star-correspondence}
\begin{gather}
\img O^{\star}{=}\imgl O^{\star}\rhoI[\img]{}^{\star}{=}\imgl O^{\star}1{=}\img{\cal W}_{\idx{dir}}(\hat O);~~~
\img\rho^{\star}{=}\img{\cal W}_{\idx{dir}}(\hat \rho_{\idx{red}});\\
\hat \rho_{\idx{red}}{=}\hat{\cal W}_{\idx{rev}}(\img\rho^{\star});~~~\hat O{=}\hat{\cal W}_{\idx{rev}}(\img O^{\star}),
\end{gather}
\end{subequations}
where the direct and reverse transforms are defined as:
\begin{gather}\label{euler.-isomorhhism_operators}
\img{\cal W}_{\idx{dir}}(\hat\odot){=}\Tr[\hat\odot\img{\hat\Delta}];~~~%\\
\hat{\cal W}_{\idx{rev}}(\img\odot){=}\left(\frac1{\hbar^2(\pi L)},\img{\hat\Delta}{\img\odot}\right)_{\idx{W}}.
\end{gather}
Here $\img{\hat\Delta}$ is the Stratonovich-Weyl (SW) kernel:
\begin{gather}\label{euler.-Stratonovich_kernel}
\img{\hat\Delta}{=}\sum_{l,k_1,k_2}\img\rho_{l,k_1;l,k_2}^{\star}\hat\rho_{l,k_2;l,k_1},
\end{gather}
and the basis functions $\hat\rho_{l,k_1;l,k_2}$ are related to ones defined by eqs.~\eqref{euler.-image_of_|l,l><l,l|-projector} and \eqref{euler.-image_of_|l,k_1><l,k_2|-projector} as:
\begin{gather}
\img\rho_{l,k_1;l,k_2}^{\star}{=}\hbar(\pi L)^{\frac12}\img\rho_{l,k_1;l,k_2}.
\end{gather}

Remarkably, the mathematical structure of the resulting Wigner images $\imgl L_k^{\star}$ is identical (up to the complex conjugation) to the conventional generalized Bopp operators for spin \cite{2007-Zueco,2010-Polkovnikov}. This analogy makes it evident that the correspondences \eqref{euler.-star-correspondence} allow to define the familiar phase space star product $\star$ of any two Weyl symbols:
\begin{gather}\label{euler.-star_product_definition}
\img\odot_1^{\star}{\star}\img\odot_2^{\star}{=}\imgl\odot_1^{\star}\img\odot_2^{\star}{=}
%\img{\cal W}_{\idx{dir}}(\hat\odot_1\hat\odot_2){=}
\img{\cal W}_{\idx{dir}}(\hat{\cal W}_{\idx{rev}}(\img\odot_1^{\star})\hat{\cal W}_{\idx{rev}}(\img\odot_2^{\star})),
\end{gather}
so that e.g.
\begin{gather}
\midop{\hat F}{=}\left(\frac1{\hbar^2(\pi L)},\imgl F^{\star}\img \rho^{\star}\right)_{\idx{W}}{=}\left(\frac1{\hbar^2(\pi L)},\img F^{\star}{\star}\img \rho^{\star}\right)_{\idx{W}}.
\end{gather}

This example shows that the proposed dynamic algorithm (\ref{intro.-def_Wigner_dyn}) does not necessarily lead to outcomes fully consistent with the SW formalism and affords additional capabilities to give the quantizers desirable properties beyond the scope of the SW framework. 
In principle, one can similarly construct the $\star$-version of any generalized Wigner quantizer. However, as we will see in Sec.~\ref{@SEC:Complete}, in general case there is no guarantee that the complete consistency with SW formalism will be achieved.

\EDITEND

3. It follows from Eq.~\eqref{euler.-[L_re,L_re]=[L_im,L_im]=L_re} that $\frac12(\imgl L_i{+}\imgr L_i){=}\Lr[i]{\ne}L_i$ regardless of the particular choice of averaging and normalization. This precludes the analogs of relations \eqref{intro.-<q^n>,<p^n>} for angular components $\img L_k$, so one can no longer obtain meaningful marginal distributions via partial integration over $\img \rho$. Nevertheless, other characteristic semiclassical features of the Wigner representation remain preserved. Specifically, one can still apply the recipe from \cite{2012-Bondar,2013-Bondar} to relate the Wigner equations of motion \eqref{intro.-Liouville_equation} for pure states (with $c_1{=}{-}3$, $\xi{=}1/2$) to the respective classical equations in standard or Koopman von Neumann form.

4. The truncated Euler phase space $\genphasespace[L]$ is incapable of handling the orientation of top relative to the laboratory frame $S'$. In particular, we can not define the $\vec e'_3$-projection of the angular momentum and the associated quantum number $m$. Also, it is easy to verify the equality $\imgl {L^2}{=}\imgr {L^2}$ which gives rise to the relation $[\hat L^2,\hat\rho_{\idx{red}}]{=}(\imgl{L^2}{-}\imgr{L^2})\img\rho{=}0$ and many-to-one ambiguity
\begin{gather}\label{euler.-many-to-one}
\forall \alpha:\alpha\hat L^2\hat{\rho}_{\idx{red}}{+}(1-\alpha)\hat{\rho}_{\idx{red}}\hat L^2~~~\LR~~\imgl{L^2}\img{\rho}.
\end{gather}\EDITBEGIN
Consequently, the feasible density matrices $\hat\rho_{\idx{red}}$ must obey the condition $\forall l_1{\ne}l_2:\Tr[\hat\rho_{l_1,k_1;l_2,k_2}\hat\rho_{\idx{red}}]{=}0$, which justifies the specific form of the isomorphism \eqref{euler.-rho-isomorphism}.

\EDITEND

Beyond that, the equality $\imgl {L^2}{=}\imgr {L^2}$ implies that the quantum Liouvillian of the free spherical top exactly coincides with its classical counterpart: $\imgl{\cal L}{=}{\cal L}{=}0$.

5. By virtue of the many-to-one ambiguousness \eqref{euler.-many-to-one} the equation $\imgl{L^2}\rhoI[\img]{l}{=}\hbar l(l{+}1)\rhoI[\img]{l}$ has bounded, isotropic, Lebesgue- and square-integrable in $\genphasespace$ solutions $\rhoI[\img]{l}$ for any real $l{>}{-}\frac12$ (see Appendix~\ref{@SEC.-app1}, Eq.~\eqref{app1.-i-solution}). However, the coefficients $\kappa_{l,j}$ in expansion $\rhoI[\hat]{l}{=}\sum_{j{=}0}^{\infty}\kappa_{l,j}\rhoI[\hat]{\frac{j}{2}}$ take negative values for non-integer values of $2l$ since $\kappa_{l,j}|_{2l{\not\in}\mathbb{Z},j{\to}\infty}{\propto}\frac{({-}1)^j}{j}$. Consequently,  $\rhoI[\hat]{l}|_{2l{\not\in}\mathbb{Z}}$ are Weyl symbols of non-positive operators and do not represent valid physical states. These properties should be considered with caution in calculations because they indicate that a small numerical error can result in a dramatic physical mistake.

\section{Complete phase space representation of rotational motion\label{@SEC:Complete}}
\begin{figure}[t]
\includegraphics[width=0.20\textwidth]
{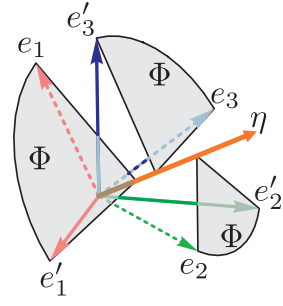}
\caption{The physical meaning of the parameters entering the definition \eqref{complete.-quaternions_def} of the quaternions $\lambda_k$.\label{*FIG.00}}
\end{figure}

Various sets of generalized coordinates enable establishing the link of the rotational dynamics with the laboratory frame missed in the Euler quantization picture. However, the evolution equations take the most elegant form in terms of the four quaternions $\lambda_k$ defined as:
\begin{gather}
\lambda_0=\cos\frac{\Phi}{2};~~~
\lambda_k{=}\eta_k\sin\frac{\Phi}{2}~~(k=1,2,3)\label{complete.-quaternions_def},
\end{gather}
where the parameters $\vec\eta$ and $\Phi$ are such that rotation about the vector $\vec\eta$ by angle ${-}\Phi$ will superimpose the axes $\vec e_k$ and $\vec {e_k}'$ of the moving and laboratory frames $S$ and $S'$ (Fig.~\ref{*FIG.00}). Unlike angular variables, the quaternions are ``true canonical coordinates'' (in the sense Ref. \cite{1999-Poirier}). This makes the construction of the Wigner representation in terms of $\lambda_i$ and the associated canonically conjugated momenta $p_{\lambda,i}$ straightforward since the associated Bopp operators obey the canonical commutation relations identical to \eqref{intro.-left_operators_def}, \eqref{intro.-right_operators_def} \cite{1969-Pierre,1970-Nienhuis} (see Appendix~\ref{@SEC.-app0} for details and brief review of the algebra of quaternions). However, extra dimensionality of the phase space makes this approach computationally impractical.

In order to solve this problem while keeping the simple form of the dynamical equations, we will consider the non-canonical phase space $\genphasespace[\lambda,L']$ composed of $\lambda_k$ and the projections $L_k'$ of the angular momentum on the laboratory axes. The corresponding classical Liouvillian reads:
\begin{gather}\label{complete.-classical_Liouvillian}
{\cal L}{=}\frac12\sum _{k{=}1}^3 \sum_{m,n{=}0}^3\mq_{m,n,k}\omega_{k}\lambda_{m} \pder{}{\lambda_n},
\end{gather}
where $\mq_{m,n,k}$ are quaternion multiplication coefficients:
\begin{gather}\label{complete.-qtn_multiplication_coeffs}
\mq_{k,i,j}{=}\begin{cases}
\epsilon_{i, j, k}&\mbox{if }i{>}0{\wedge}j{>}0{\wedge}k{>}0;\\
\delta_{j,0}\delta_{k,i}{+}\delta_{i, 0}\delta_{k,j}{-}\delta_{k,0}\delta_{i,j}& \mbox{otherwise},
\end{cases}
\end{gather}
and $\omega_k{=}{\sum_{j{=}1}^{3}Q_{k,j}L_j'}/{I_k}$ are the angular frequencies about axes $\vec e_k$. The entries $Q_{i,j}$ of the directional cosine matrix are bilinear in terms of $\lambda_k$: $Q_{i,j}{=}(\vec e_{i},\vec e'_{j})=\sum_{m,n}q_{i,j,m,n}\lambda_m\lambda_n$, with coefficients
\begin{gather}\label{complete.-directional_cosines_coeffs}
q_{i,j,m,n}{=}(1{-}2\delta_{j,m})\sum_{k{=}0}^{3}\mq_{i,j,k}\mq_{k,m,n}.
\end{gather}

\EDITBEGIN

In choosing a strategy to quantize  Eqs.~\eqref{complete.-quaternions_def} and \eqref{complete.-classical_Liouvillian}, we will follow the reasoning of the previous section and
define a set of postulates similar to (E:\ref{euler.-postulate_P-real}-\ref{euler.-postulate_cl_limit}):
\renewcommand{\labelenumi}{(C:\arabic{enumi})}
\noindent\begin{enumerate}\itemsep1pt \parskip0pt \parsep0pt
\item \label{complete.-postulate_P-real} the enforced reality condition similar to (E:\ref{euler.-postulate_P-real});
\item \label{complete.-postulate_(P_1,P_2)=0} the traciality relation \eqref{intro.-(P_1,P_2)=0} (with $(\img\odot_1,\img\odot_2)_{\idx{W}}\overset{\idx{def}}{=}\int_{\genphasespace[\lambda,L']}\img\odot_1^*\img\odot_2dL_1dL_2dL_3d\lambda_0d\lambda_1d\lambda_2d\lambda_3$); %$d\Omega{=}\prod_{i=1}^3dL_i\prod_{j=0}^3d\lambda_j$);
\item \label{complete.-postulate_cl_limit} the proper classical limits:
\begin{gather}\label{complete.-classical_limits}
\imgl L_i'|_{h{\to}0}{=}L_i';~\imgl L_i|_{h{\to}0}{=}L_i;~\imgl{\cal L}|_{\hbar{\to}0}{=}{\cal L};~\imgl{\lambda}_j|_{h{\to}0}{=}\lambda_i,
\end{gather}
\end{enumerate}
together with the commutation relations \eqref{euler.-[L_i,l_j]}-\eqref{euler.-[L_re,L_im]} and
\begin{gather}
\frac{2}{i\hbar}[\hat L_i,\hat{\lambda}_j]{=}\sum_{k{=}1}^3\mq_{k,j,i}\hat{\lambda}_k.
\end{gather}
As in Sec.~\ref{@SEC:Euler}, our intention is to make use of the variability in the framework of above postulates in favor of the simplest form of the phase space Liouville operator $\imgl{\cal L}$.

\EDITEND

Reproducing the steps leading to Eqs.~\eqref{euler.-L_im} and \eqref{euler.-L_re} one gets:
\begin{gather}
\imgl{L}_k{=}\frac12\left[\frac{\sum_{s{=}1}^3Q_{k,s}L_s'}{\sum_{s{=}0}^3{\lambda_s^2}},1{-}\frac{\hbar^2}{16}\sum_{s{=}1}^3\pder{^2}{L_s'^2}\right]_+{+}\notag\\
\sum_{m,n{=}0}^3\left(\frac{\hbar^2}{16}\sum_{s{=}1}^3q_{k,s,m,n}\pder{}{L_s'}{+}i\frac{\hbar}{4}\mq_{m,n,k}\right)\lambda_m\pder{}{\lambda_n};\label{complete.-A:L_k}\\
\imgl{\lambda_k}{=}\imgl{N}^{{-}\frac12}\left(\lambda_k{+}i\frac{\hbar}{4}\sum_{m={0}}^3\sum_{n{=}1}^3\mq_{k,n,m}\lambda_m\pder{}{L_n'}\right);\label{complete.-A:lambda_k}\\
\imgl{N}{=}\left(\sum_{s{=}0}^3\lambda_s^2\right)\left(1{-}\frac{\hbar^2}{16} \sum_{s{=}1}^3\pder{^2}{L_s'^2}\right)\label{complete.-A:N},
\end{gather}
where the operator $\imgl{N}$ commutes with all physical observables of the form $F(\imgl{L'}_k,\imgl{\lambda_n})$. The existence of such $\imgl{N}{\ne}1$
is due to overcompleteness of our 7D phase space $\genphasespace[\lambda,L']$.

The Wigner representation \eqref{complete.-A:L_k}-\eqref{complete.-A:N} is not convenient for exact numerical implementation. One problem is caused by the term $\imgl{N}^{{-}\frac12}$ in the expression \eqref{complete.-A:lambda_k} for the quaternion images, which is a differential operator of infinite order. This pre-factor, however, can be ignored when choosing to work only with images of states satisfying equation $\imgl N\img{\rho}{=}\img{\rho}$ (which is possible owing to its commutation properties). Another possibility is to Fourier or Laplace transform the phase space $\genphasespace[\lambda,L']$ with respect to $L_1'$, $L_2'$ and $L_3'$. However, the resulting equations will loose the key characteristic properties of the Wigner representation. Another technical complication is introduced by the excessive dimensionality of $\genphasespace[\lambda,L']$. In addition, unlike the classical generator of free motion \eqref{complete.-classical_Liouvillian} the quantum counterpart \eqref{intro.-Liouvillian} no longer evidently manifests the angular momentum conservation by preserving the values of $L_k'$. One would desire to retain this remarkable property of Eq.~\eqref{complete.-classical_Liouvillian} in the quantum case because it would allow reduction of the 7-dimensional differential propagation equation to a series of 4-dimensional ones.

\EDITBEGIN

We found that these issues can be resolved by relaxing the traciality requirement (C:\ref{complete.-postulate_(P_1,P_2)=0}) in favor of more explicit specification of the desired form of the Bopp operators $\imgl L_k$. The resulting modified set of postulates imposes:

\EDITEND
\renewcommand{\labelenumi}{(\={C}:\arabic{enumi})}
\noindent\begin{enumerate}\itemsep1pt \parskip0pt \parsep0pt
\item \label{complete.-postulate_rev_F-real} the reality condition \eqref{euler.-right_operators_def};
\item \label{complete.-postulate_rev_cl_limit} the requirement \eqref{complete.-classical_limits} of well-defined classical limits;
\item \label{complete.-postulate_rev_no_L'_derivatives} the requirement of the absence of derivatives over $L_1'$, $L_2'$ and $L_3'$ in the expressions for the images $\imgl L_k$.
\end{enumerate}

The expressions for $\imgl L_k$ satisfying these postulates can be compactly written in terms of the new variables $\Lambda_m{=}\sqrt{L'}\sqrt{\frac{8}{\hbar }}\lambda_m$:
\begin{gather}
\imgl{L}_k{=}\frac{\hbar}{8}\sum_{m,n{=}0}^3\left(
\sum_{s{=}1}^3
q_{k,s,m,n}\frac{L_s'}{L'}\left(\Lambda_m\Lambda_n{-}\pder{^2}{\Lambda_m\partial\Lambda_n}\right)\right.\notag\\
\left.{+}2\,i\,\mq_{m,n,k}\Lambda_m\pder{}{\Lambda_n}\right),\label{complete.-B:L_k}
\end{gather}
where $L'{=}\sqrt{\sum_{k{=}1}^3L_k'^2}$. It is useful to introduce the intermediate fixed frame $S''$ whose third axis $\vec e_3''$ coincides with the (conserved) direction of the angular moment. We  denote by $\frac{1}{L'}\set\qL$ the quaternion which represents the rotation connecting $S'$ and $S''$ and introduce the parameters $\xLambda_m$ as  exact analogs of the parameters $\Lambda_m$ characterizing the orientation of the rotor relative to $S''$:
\begin{gather}\label{complete.-(L,Lambda)->(sampi,qL)-transform}
L_n'{=}\sum_{m,n=0}^3q_{i,j,m,n}\qL_i\qL_j;~\Lambda_k{=}\frac{\sum_{i,j{=}0}^3\mq_{k,i,j}\qL_i\xLambda_j}{\sqrt{\sum_{n=0}^3\qL_n^2}}.
\end{gather}
Note that Eqs.~\eqref{complete.-(L,Lambda)->(sampi,qL)-transform} do not fix the directions of the axes $\vec e_1''$ or $\vec e_2''$ of $S''$, and so do  not uniquely define $\qL_k$.
Eqs.~\eqref{complete.-B:L_k} take simple forms in terms of the new variables $\xLambda_m$:
\begin{gather}
\imgl L_1{\pm}i\imgl L_2{=}\hbar\imgl a_{\mp}^{\dagger}\imgl a_{\pm};~~~
\imgl L_3{=}\frac{\hbar}2(\imgl a_{+}^{\dagger}\imgl a_{+}{-}\hat a_{-}^{\dagger}\hat a_{-})\label{complete.-B:L_Schwinger_oscillator},
\end{gather}
where $\imgl a^{\dagger}_{\pm}$ and $\imgl a_{\pm}$ are the conventional ladder operators:
\begin{gather}
{\imgl a^{\dagger}_{\pm}{+}\imgl a_{\pm}}{=}
{\xLambda_{2{\pm}1}{\pm}i\pder{}{\xLambda_{1{\mp}1}}};~
{\imgl a_{\pm}{-}\imgl a^{\dagger}_{\pm}}{=}
\pder{}{\xLambda_{2{\pm}1}}{\pm}i\xLambda_{1{\mp}1};\label{complete.-B:a-ladder_operators}\\
[\imgl a_{\pm},\imgl a^{\dagger}_{\pm}]{=}1;~~~[\imgl a_{\mp},\imgl a^{\dagger}_{\pm}]{=}[\imgl a_{\mp},\imgl a_{\pm}]{=}0.\label{complete.-B:a-ladder_commutators}
\end{gather}
(It is worth stressing that notations such as $\imgl a^{\dagger}_{\pm}$ hereafter mean ``the Wigner image of the creation operator $\hat a^{\dagger}$'', not the ``Hermitian conjugate of the Bopp operator $\imgl a_{\pm}$'').

The fact that $\imgl a^{\dagger}_{\pm}$ and $\imgl a_{\pm}$ do not depend on $\qL_k$ and the commutation relation $\forall m,n:[\hat L_m,\hat L_n']{=}0$ hints that the images $\imgl L_m'$ should have a form similar to \eqref{complete.-B:L_Schwinger_oscillator}:
\begin{gather}
\imgl L_1'{\pm}i\imgl L_2'{=}\hbar\imgl b_{\pm}^{\dagger}\imgl b_{\mp};~~~
\imgl L_3'{=}\frac{\hbar}2(\imgl b_{+}^{\dagger}\imgl b_{+}{-}\hat b_{-}^{\dagger}\hat b_{-})\label{complete.-B:L'_Schwinger_oscillator},
\end{gather}
where the operators  $\imgl b^{\dagger}_{\pm}$ and $\imgl b_{\pm}$ do not depend on $\xLambda_k$ and satisfy  commutation relations identical to \eqref{complete.-B:a-ladder_commutators}. The validity of Eqs.~\eqref{complete.-B:L'_Schwinger_oscillator} is proven in Appendix~\ref{@SEC.-app2}, where the following explicit expressions are obtained (up to invariance transformation \eqref{complete.-B:invariance_relation}, see below):
\begin{gather}
\hspace{-6pt}\imgl b_{\pm}^{\dagger}{=}\sqrt2(\qL_{1{\mp}1}{+}i\qL_{2{\pm}1});
~
\imgl b_{\pm}{=}\frac{1}{\sqrt8}(\pder{}{\qL_{1{\mp}1}}{-}i\pder{}{\qL_{2{\pm}1}});\label{complete.-B:b-ladder_operators}\\
\imgl L_k'{=}\frac{L_k'}{L'}
\frac{\imgl\Lqn{+}\imgr\Lqn}{2}{-}
\frac{\hbar}{4}\sum_{r{=}0}^3\sum_{s{=}1}^3\mq_{k,r,s}((1{-}\delta_{r,0})\frac{L_r'}{L'}{+}i\delta_{r,0}){\times}{}\notag\\
\sum_{m,n{=}0}^{3}\mq_{n,s,m}\left({2(1{-}\delta_{m{\times}n,0})L_m'\pder{}{L_n'}{+}\Lambda_m\pder{}{\Lambda_n}}\right),\label{complete.-B:L_k'}
\end{gather}
where $\imgl \Lqn$ is the image of operator of quantum number $l$:
\begin{gather}
\imgl \Lqn(\imgl \Lqn{+}\hbar){=}\sum_{k{=}1}^3\imgl L_k^2{=}\sum_{k{=}1}^3\imgl {L_k'}^2{=}{\imgl L^2};\label{complete.-B:L(Schrodinger)}\\
\imgl \Lqn{=}\frac{\hbar}8\sum_{m{=}0}^{3}(\Lambda_m^2{-}\pder{^2}{\Lambda_m^2}{-}1{-}2i\sum_{k{=}1}^3\sum_{n{=}0}^3\mq_{n,k,m}\frac{L_k'}{L'}\Lambda_m\pder{}{\Lambda_n})\label{complete.-B:L}.
\end{gather}

 We emphasize several important properties of the Wigner quantizer generated by operators \eqref{complete.-B:L_k} and \eqref{complete.-B:L_k'}:

\EDITBEGIN

1. The following invariance relations hold for any operator $\cal S$ representable as a function of only $\imgl \Lqn$, $\imgr \Lqn$, $L'$ and the operator $\imglr Q{=}\sum_{s{=}1}^{3}L'_s\pder{}{L'_s}$:
\begin{gather}\label{complete.-B:invariance_relation}
\imgl L_k{=}{\cal S}\imgl L_k{\cal S}^{-1};~~~\imgl L_k'{=}{\cal S}\imgl L_k'{\cal S}^{-1}.
\end{gather}
In particular, if $\img \rho_{\epsilon}$ is the solution of the eigenvalue problem $f(\imgl L_k,\imgl L_l',\imgr L_m,\imgr L_n',L')\rho_{\epsilon}{=}\epsilon \rho_{\epsilon}$ with arbitrary function $f$, then ${\cal S}\rho_{\epsilon}$ is also its solution. Furthermore, the variables replacement $L_k'{\to}\epsilon L_k'$ with arbitrary $\epsilon$ does not change the form of the operators \eqref{complete.-B:L_k'}, \eqref{complete.-B:L_k} and  \eqref{complete.-B:L}. This implies that the basis function $\img\rho_{\alpha,\beta}$ corresponding to an arbitrary projector $\hat\rho_{\alpha,\beta}{=}\ket{l_{\alpha},m_{\alpha},k_{\alpha}}\bra{l_{\beta},m_{\beta},k_{\beta}}$ can be written as:
\begin{gather}\label{complete.-B:rho-6D}
\img\rho_{\alpha,\beta}{=}r_{L',l_{\alpha},l_{\beta}}\img\rho^{(0)}_{\alpha,\beta}(\set\Lambda,\frac{L_1'}{L'},\frac{L_2'}{L'},\frac{L_3'}{L'}),
\end{gather}
where the variable prefactor $\imglr r_l(L')$ depends on the choice of $\cal S$ in the invariance relation \eqref{complete.-B:invariance_relation}.

2. The Bopp operators $\imgl L_k$, $\imgr {L^2}$, and
$\frac12(\imgl L_3'{-}\imgr L_3')$ are Hermitian in $\genphasespace[\Lambda,L']$, but the operators $\imgl L_k'$ are not:
\begin{gather}\label{complete.-B:L_k'^dagger}
\imgl L_k'^{\dagger}{=}\imgl L_k^*|_{\pder{}{\Lambda_n}{\to}{-}\pder{}{\Lambda_n},\pder{}{L_m}{\to}{-}\pder{}{L_m}}{+}\hbar\frac{L_k'}{L'}{\ne}\imgl L_k'.
 \end{gather}
(The extra term $\hbar\frac{L_k'}{L'}$ in \eqref{complete.-B:L_k'^dagger} arises from  symmetrization of the operators \eqref{complete.-B:L_k'}.) The Bopp operators $\imgl L_k'^{\dagger}$ also fulfill relations \eqref{euler.-[L_i,l_j]},  \ref{complete.-B:L(Schrodinger)} and can be used as an alternative variant of the images of the operators $\hat L_k'$. Nevertheless, the original isomorphism $\hat L_k'{\LR}\imgl L_k'$ results in more convenient forms of the Weyl symbols. Indeed, denote as $\rhoI[\img]{l}$ the Weyl symbols of the identity submatrices $\sum_{k,m{=}{-}l}^l\ket{l,m,k}\bra{l,m,k}$ for subspaces with well-defined quantum number $l$. Each of $\rhoI[\img]{l}$ should be the symmetric solution of the eigenvalue problem $\imgl \Lqn~\rhoI[\img]{l}{=}l\rhoI[\img]{l}$. If the isomorphism $\hat L_k'\LR\imgl L_k'$ is accepted, one obtains:
\begin{gather}\label{app3.-rhoI}
\rhoI[\img]{l}{=}r_{l}(L')\Laguerre_{2l}^{1}\farg{\sum_{m{=}0}^{3}\Lambda_m^2}e^{{-}\frac12\sum_{m{=}0}^{3}\Lambda_m^2},
\end{gather}
and $\rhoI[\img]{l}|_{\Lambda_m{\to}{\pm}\infty}{=}0$. Conversely, the choice $\hat L_k'{\LR}\imgl L_k'^{\dagger}$ leads to the divergent solutions $\rhoI[\img]{l}{\propto}\Laguerre_{2l}^{1}\farg{\sum_{m{=}0}^{3}\Lambda_m^2}e^{{+}\frac12\sum_{m{=}0}^{3}\Lambda_m^2}$.
This substantially complicates the definition of normalization and the rule for calculation of averages and makes this choice inconvenient. However, even the original definition \eqref{complete.-B:L_k} leads to non-orthogonality of certain basis functions because of the non-Hermiticity of the operators $\imgl L_k'$:
\begin{gather}
\iiint\prod_{k{=}1}^3dL_k\iiiint\prod_{m{=}0}^3d\Lambda_k\rho_{\alpha_1,\beta_1}^*\rho_{\alpha_2,\beta_2}{\ne}0\notag
\end{gather}
for any pair of basis functions such that $l_{\alpha_1}{=}l_{\alpha_2}$, $l_{\beta_1}{=}l_{\beta_2}$, $k_{\alpha_1}{=}k_{\alpha_2}$, $k_{\beta_1}{=}k_{\beta_2}$ $m_{\alpha_1}{-}m_{\alpha_2}{=}m_{\beta_1}{-}m_{\beta_2}$, including the cases where $m_{\alpha_1}{\ne}m_{\alpha_2}$. Thus, the discussed Wigner representation can not be equipped with a traciality relation similar to \eqref{intro.-(P_1,P_2)=0}.

Nevertheless, thanks to the invariance relation \eqref{complete.-B:invariance_relation}, one can define a convenient rule for the calculation of averages by selecting the following $L'$-independent prefactor $r_{L',l_{\alpha},l_{\beta}}$ in eq. \eqref{complete.-B:rho-6D}:
\begin{gather}\label{app4.-r_l}
r_{l_{\alpha},l_{\beta}}{=}\frac{(-1)^{l_{\alpha}{+}l_{\beta}}}{16 \pi^3}\sqrt{2l_{\alpha}{+}1}\sqrt{2l_{\beta}{+}1}.
\end{gather}
In this case only 6 out of the 7 arguments of the Wigner function $\img\rho{=}\img\rho(\set\Lambda,\frac{L_1'}{L'},\frac{L_2'}{L'},\frac{L_3'}{L'})$ are independent, so that the effective size of the phase space is equal to 6. Furthermore, the averaging rule is defined through the scalar product \eqref{intro.-(F,P)}, where
\begin{gather}\label{complete.-B:(F,P)}
( \odot_1^*,\odot_2)_{\idx{W}}{=}\iiiint_{-\infty}^{\infty}d\Lambda_0d\Lambda_1d\Lambda_2d\Lambda_3\int_{\Omega}\set{d\omega}\frac{\odot_1\odot_2}{\hbar^2\kappa^2}.
\end{gather}
The inner integral $\int_{\Omega}\set{d\omega}...$ in \eqref{complete.-B:(F,P)} is taken over the surface of a  sphere $L'{=}\kappa\hbar$ with an arbitrary radius $\kappa$.

3. Lack of a traciality relation like (C:\ref{complete.-postulate_(P_1,P_2)=0}) makes elucidation of the correspondence rule $\hat\rho{\LR}\img\rho$ less straightforward than in the canonical case. Formally, the Weyl symbol $\img\rho_{\alpha,\beta}$ of any basis function $\hat\rho_{\alpha,\beta}$ can be obtained by sequential application of the Schwinger ladder operators \eqref{complete.-B:a-ladder_operators} and \eqref{complete.-B:b-ladder_operators} to the ground state $\rho{=}\rhoI[\img]{0}$. However, the direct analogs of these operators can not exist in $\genphasespace[\Lambda,L']$ space because they would result in two independent and conflicting definitions for the operator $\imgl{L^2}$. Nevertheless, the Wigner images $\imgl{a_{\pm} b_{\pm}}$ and $\imgl{a^{\dagger}_{\pm}b^{\dagger}_{\pm}}$ of the compound Schwinger operators $\hat a_{\pm}\hat b_{\pm}$ and $\hat a_{\pm}^{\dagger}\hat b_{\pm}^{\dagger}$ are well-defined and can be directly deduced by the technique used in Appendix~\ref{@SEC.-app2}. The derivation and resulting rather cumbersome expressions are deferred to Appendix~\ref{@SEC.-app4}.

The compound ladder operators allow to explicitly calculate any basis function \eqref{complete.-B:rho-6D}:
\begin{gather}\label{complete.-projector-explicit}
\img\rho_{\alpha,\beta}{=}\imglr {\cal R}_{\uparrow}(\alpha,\beta)\rhoI[\img]{0};~~~\rhoI[\img]{0}{=}\imglr {\cal R}_{\downarrow}(\alpha,\beta)\img\rho_{\alpha,\beta},
\end{gather}
where
\begin{gather}
\imglr {\cal R}_{\uparrow}(\alpha,\beta){=}\prod_{\kappa,\mu{=}\pm1}
\frac{
\left(\imgl{a^{\dagger}_{\kappa}b^{\dagger}_{\mu}}\right)^{p_{\kappa,\mu}(\alpha)}\left(\imgr{a^{\dagger}_{\kappa}b^{\dagger}_{\mu}}\right)^{p_{\kappa,\mu}(\beta)}
}%
{\prod_{\lambda{=}m,k}\prod_{\xi{=}\alpha,\beta}\sqrt{(l_{\xi}{-}\lambda_{\xi})!(l_{\xi}{+}\lambda_{\xi})!}};\\
\imglr {\cal R}_{\downarrow}(\alpha,\beta){=}\prod_{\kappa,\mu{=}\pm1}
\frac{
\left(\imgl{a_{\kappa}b_{\mu}}\right)^{p_{\kappa,\mu}(\alpha)}\left(\imgr{a_{\kappa}b_{\mu}}\right)^{p_{\kappa,\mu}(\beta)}
}%
{\prod_{\lambda{=}m,k}\prod_{\xi{=}\alpha,\beta}\sqrt{(l_{\xi}{-}\lambda_{\xi})!(l_{\xi}{+}\lambda_{\xi})!}},
\end{gather}
and the factors $p_{\kappa,\mu}$ can be any set of nonnegative numbers satisfying the relations:
$\sum_{\kappa,\mu{=}\pm1}p_{\kappa,\mu}(\xi){=}2l_{\xi}$, $\sum_{\kappa,\mu{=}\pm1}\kappa p_{\kappa,\mu}(\xi){=}2 k_{\xi}$, $\sum_{\kappa,\mu{=}\pm1}\mu p_{\kappa,\mu}(\xi){=}2 m_{\xi}$. Using the operator $\imglr {\cal R}_{\downarrow}$ and the orthogonality relation $\forall\img\rho_{\alpha,\beta}{\ne}\rhoI[\img]{0}:(\rhoI[\img]{0},\img\rho_{\alpha,\beta})_{\idx{W}}{=}0$ (the latter follows from the Hermiticity of the images $\imgl L_k$) one can establish the desired correspondences $\img\rho{=}\img{\cal W}_{\idx{dir}}(\hat \rho)$ and
$\hat \rho{=}\hat{\cal W}_{\idx{rev}}(\img\rho)$:
\begin{subequations}\label{complete.-B:isomorhhism}
\begin{gather}
\img{\cal W}_{\idx{dir}}(\hat\rho){=}\sum_{\alpha,\beta}\imglr {\cal R}_{\uparrow}(\alpha,\beta)\rhoI[\img]{0}\Tr[\hat\rho_{\alpha,\beta}^{\dagger}\hat\rho];\label{complete.-B:isomorhhism_dir}\\
\hat{\cal W}_{\idx{rev}}(\img\rho){=}\sum_{\alpha,\beta}(\rhoI[\img]{0},\imglr {\cal R}_{\downarrow}(\alpha,\beta)\img\rho)_{\idx{W}}\rho_{\alpha,\beta}{=}\notag\\
\sum_{\alpha,\beta}(\imglr {\cal R}_{\downarrow}^\dagger(\alpha,\beta)\rhoI[\img]{0},\img\rho)_{\idx{W}}\hat\rho_{\alpha,\beta}.\label{complete.-B:isomorhhism_rev}
\end{gather}
\end{subequations}
It is worth stressing that the notation $\imglr {\cal R}_{\downarrow}^\dagger(\alpha,\beta)$ means ``the Hermite conjugate of the phase space Bopp operator $\imglr {\cal R}_{\downarrow}(\alpha,\beta)$ in $\genphasespace[\Lambda,L']$'' rather than ``the Wigner image of the Hermite conjugate $\hat {\cal R}_{\downarrow}^{\dagger}$ of the associated operator $\hat {\cal R}_{\downarrow}$''.

4. Unlike the Wigner quantization of the Euler equations, the traciality-deficient $\{\Lambda,L'\}$-representation can not be tuned to become fully consistent with the standard Stratonovich-Weyl quantization scheme. Specifically, it is still possible to achieve the identity $\rhoI[\img]{}^{\star}{=}\rhoI[\imgl]{}^{\star}{=}1$ by applying the suitable invariance transform \eqref{complete.-B:invariance_relation} with ${\cal S}{=}{\cal S}^{\star}{=}\frac{256\pi^3}{{\sum_{m{=}0}^{3}}\Lambda_m^2}$:
\begin{gather}
\img\rho^{\star}{=}{\cal S}^{\star}\img\rho;~~~\imgl F^{\star}{=}S^{\star}\imgl F ({\cal S}^{\star})^{{-}1},
\end{gather}
and introduce the Weyl symbols of operators: $\img F^{\star}{=}\imgl F^{\star}\rhoI[\img]{}^{\star}$. However, the inequality $\imglr {\cal R}_{\uparrow}^{\star}(\alpha,\beta){\ne}{\imglr {\cal R}_{\downarrow}^{\star}}^{\dagger}(\alpha,\beta)$ makes it impossible to define the Stratonovich-Weyl kernel similar to \eqref{euler.-Stratonovich_kernel}. Instead, one has to introduce the direct and reverse transforms $\img{\cal W}_{\idx{dir}}^{\star}(\hat\odot){=}(S^{\star}\img{\cal W}_{\idx{dir}}(\hat\odot)$,
$\hat{\cal W}_{\idx{rev}}^{\star}(\img\odot){=}\hat{\cal W}_{\idx{rev}}((S^{\star})^{{-}1}\img\odot)$ as independent operations. These transforms nevertheless allow to define the analog of the star product algebra similar to \eqref{euler.-star_product_definition}, so that e.g.:
\begin{gather}
\midop{\hat F}{=}({\cal S}^{-1},\imgl{F}\img{\rho})_{\idx{W}}{=}({\cal S}^{-1},\img{F}{\star}\img{\rho})_{\idx{W}}.
\end{gather}

\EDITEND

5. None of the expressions $\frac12(\imgl L_k'{+}\imgr L_k')$ and  $\frac12(\imgl \lambda_n'{+}\imgr \lambda_n')$ coincides with its classical analog $L_k$, $L_k'$ and $\lambda_n$, hence the marginal distributions associated with the Wigner function $\img\rho$ have no exact physical meaning. Instead, any of the representations $\{\Lambda,L'\}$, $\{\Lambda,q\}$, and $\{\xLambda,\qL\}$ allow to easily cast the generators of motion $\imgl {\cal L}$ for free  linear, spherical or symmetric tops with principal moments of inertia $I_1{=}I_2{\ne}I_3$ in the familiar classical-like form:
\begin{gather}\label{complete.-B:symmetric_top_Liouvillian}
\imgl{\cal L}{=}{-}A\imglr\Jqn\pder{}{\alpha}{+}(A-B)\imglr\Kqn\pder{}{\gamma},
\end{gather}
where $A{=}\frac{1}{I_1}{=}\frac{1}{I_2}$, $B{=}\frac{1}{I_3}$ ($B{=}\imglr\Kqn{=}0$ for linear tops) and
\begin{gather}\label{complete.-B:symm_top-Jqn,Mqn-operators}
\imglr\Jqn{=}\frac{\imgl \Lqn{+}\imgr \Lqn{+}\hbar}{2};~~~
\imglr\Kqn{=}\frac{\imgl L_3{+}\imgr L_3}2,
\end{gather}
by expressing the parameters $\xLambda_m$ in terms of the Euler angles $\alpha$, $\beta$ and $\gamma$ relating the frames $S'$ and $S''$ \cite{BOOK-Borisov}:
\begin{gather}
\xLambda_0{\pm}i\xLambda_3{=}\sqrt{\frac{8L'}{\hbar}}\cos\farg{\frac{\beta}2} \exp\farg{{\pm}i\frac{\alpha {+}\gamma }{2}};~\notag\\
\xLambda_1{\pm}i\xLambda_2{=}\sqrt{\frac{8L'}{\hbar}}\sin\farg{\frac{\beta}2} \exp\farg{{\pm}i\frac{\alpha {-}\gamma }{2}}.
\end{gather}
\noindent (Note that in the definitions of Euler angles we adopted the conventions of Zare and Edmonds books \cite{BOOK-Zare,BOOK-Edmonds}.)

The only difference between the quantum Liouvillian ~\eqref{complete.-B:symmetric_top_Liouvillian} and its classical counterpart is that the classical variables $L'$ and $L_3'$ are replaced with the quantum Bopp operators $\imglr\Jqn$ and $\imglr\Kqn$ with discrete spectra. That is, the values of axial and precession frequencies of the quantum top can take only a discrete set of equidistant values $(A-B)\frac{\hbar}2k$ and $A\frac{\hbar}2(|j|+1)$ $(k,j\in\mathbb{Z})$. which gives raise to the familiar phenomenon of quantum rotational revivals.

The analogy to the classical case can be pushed even further by forcing the operators $\imgl \Lqn$ and $\imgl L_3$ to take the mathematical structure of canonical Bopp operators $\imgl x_j$ and $\imgl p_j$ (Eq.~\eqref{intro.-left_operators_def}) via an appropriate variable transformation. By comparing the form \eqref{complete.-B:symm_top-Jqn,Mqn-operators} of the operators $\imglr\Jqn$ and $\imglr\Kqn$ with the fact that  $\frac{\imgl x_j{+}\imgr x_j}{2}{=}x_j$, we can expect that such a transformation will lead us to Nasyrov-type Wigner quaintizer \cite{1999-Nasyrov}, in which the quantum generator of motion \eqref{complete.-B:symmetric_top_Liouvillian} is identical to the classical one: $\imgl {\cal L}{=}{\cal L}$. The derivation and the properties of this representation are detailed in Appendix~\ref{@SEC.-app5}. Its existence leads to the remarkable and intriguing conclusion that the free symmetric top shares with the free particle and harmonic oscillator the exceptional property of having identical classical and quantum dynamics.

\section{Summary and conclusion\label{@SEC.-concl}}

The main practical outcome of this work is the new phase space quantizers of rotation motion having a superior combination of attractive properties. Specifically, the truncated quantizer derived in Sec.~\ref{@SEC:Euler} allows one to perform the density-matrix-type calculations within the wavefunction-sized rotational phase space of 3 parameters and to fully account for any rotational effects in isotropic environments on the intramolecular dynamics. Thus, it may be useful, for instance, in calculations of  emission spectra or dissociation rates resulting from  pulsed laser excitation. We have also showed that there exists a large family of quantizers (parameterized by $c_1$ and $\xi$), including the two variants which are especially convenient for calculation of averages or the normalization of quasiprobability distributions, and the version fully consistent with the Stratonovich-Weyl formalism. One can easily switch between these representations via the simple non-unitary transformation developed.

Along with its practical potential,  this quantizer also has the pedagogical value of establishing the bridge between the formal quantization of the spin degrees of freedom \cite{2010-Polkovnikov} and the classical Euler equations.

The second proposed $\{\Lambda,L'\}$ quantizer has the important feature of translating the angular momentum conservation laws into conservation of the  parameters $L'_1$, $L'_2$, $L'_3$ in the course of free rotations. This feature allows for natural parallelization of the code via splitting the initial 6-dimensional problem into series of independent 4-dimensional ones for evolution of the parameters $\Lambda_m$ ($m{=}0,...,3$). Note that several known representations (e.g.~\cite{2013-Fischer,1999-Nasyrov})  allow a similar trick. However, their generalized parametric spaces are not singularity-free and suffer from the gimbal lock problem. The latter problem can be resolved in the framework of the standard Wigner quantization procedure only by introducing artificial degrees of freedom \cite{1969-Pierre,1970-Nienhuis}. In contrast, both of the proposed quantizers resolve the gimbal lock issue without paying this price (we recall that the $\{\Lambda,L'\}$ quantizer with scalar product \eqref{complete.-B:(F,P)} is effectively 6-dimensional). That is, they allow convenient and low-dimensional grid discretizations in numerical dynamical simulations. In addition, they benefit from expressing the generators of free motion as low-order differential operators of continuous arguments. This should facilitate relatively inexpensive propagation of the evolution equations and is the important prerequisite for effective application of the initial value approximations.

On the conceptual level, our findings uncover the direct connection (Eq.~\eqref{complete.-B:L_Schwinger_oscillator}) between the quaternion parameters and the raising and lowering operators entering the Schwinger oscillator model. This connection clarifies the physical meaning and the nature of mathematical beauty of this model.

We also established the relationship between the $\{\Lambda,L'\}$ quantizer and the Nasyrov representation \cite{1999-Nasyrov}. The latter formally allows one to reduce the quantum Liouville equation for free linear and symmetric tops to the form identical to the classical Liouville equation and propagate it using the familiar method of characteristics. In addition, to the best of our knowledge, we presented for the first time the exact differential expressions for the key Bopp operators in this representation (Appendix~\ref{@SEC.-app5}).

We hope that all the mentioned advantages will make the proposed representations useful for analysis of future experiments in quantum physics and quantum chemistry involving the complex semiclassical rotational dynamics of polyatomic molecules. We also hope that the presented results support the key conceptual proposal on the critical revision of the axiomatic approach to the formal definition of the Wigner function from the dynamical perspective. For example, we have illustrated that the $\{\Lambda,L'\}$ quantizer can not be derived within the standard Stratonovich-Weyl quantization framework. We believe that the revised axiomatization will help to achieve the desired balance between numerical utility and physical transparency when constructing the Wigner representations of other dynamical systems with nontrivial structures of the underlying phase spaces.

\acknowledgments
 We are grateful to the Department of Energy (grant number DE-FG02-04ER15612) for support and to  Dr.~Denys~I.~Bondar and Dr.~Renan Cabrera for numerous stimulating discussions, important comments and valuable suggestions.

%%%%%%%%%%%%%%%%%%%%%%%%%%%%%%%%%%%%%%%%%%%%%%%%%%%%%%%%%%%%%%%%%%%
%%%%%%%%%%%%%%%%%%%%%%%%%%%%%%%%%%%%%%%%%%%%%%%%%%%%%%%%%%%%%%%%%%%
%%%%%%%%%%%%%%%%%%%%%%%%%%%%%%%%%%%%%%%%%%%%%%%%%%%%%%%%%%%%%%%%%%%
%%%%%%%%%%%%%%%%%%%%%%%%%%%%%%%%%%%%%%%%%%%%%%%%%%%%%%%%%%%%%%%%%%%
%
%                       A P P E N D I C E S                       %
%
%%%%%%%%%%%%%%%%%%%%%%%%%%%%%%%%%%%%%%%%%%%%%%%%%%%%%%%%%%%%%%%%%%%
%%%%%%%%%%%%%%%%%%%%%%%%%%%%%%%%%%%%%%%%%%%%%%%%%%%%%%%%%%%%%%%%%%%
%%%%%%%%%%%%%%%%%%%%%%%%%%%%%%%%%%%%%%%%%%%%%%%%%%%%%%%%%%%%%%%%%%%
%%%%%%%%%%%%%%%%%%%%%%%%%%%%%%%%%%%%%%%%%%%%%%%%%%%%%%%%%%%%%%%%%%%
\appendix

\section{Finding the images of angular eigenstates in \texorpdfstring{$\genphasespace[L]{}$}{Lg}\label{@SEC.-app1} for the case \texorpdfstring{$c_1{=}{-}3$}{Lg} in \texorpdfstring{\eqref{euler.-L_re}}{Lg}
}
Consider the orthogonality relation \eqref{intro.-(P_1,P_2)=0} for the set of Weyl symbols of operators $\rhoI[\hat]{l}$:
\begin{gather}\label{app1.-[r_1,r_2]=0}
(\rhoI[\img]{l_1},\rhoI[\img]{l_2})_W{\propto}\delta_{l_1,l_2}.
\end{gather}
The Weyl symbols $\rhoI[\img]{l}$ can only depend on the scalar argument $L{=}\sqrt{L_1^2{+}L_2^2{+}L_3^2}$ due to isotropy of the operators $\rhoI[\hat]{l}$ and must be the solutions of eigenvalue problem:
\begin{gather}\label{app1.-L^2r=hl(l+1)r}
\imgl{L^2}\rhoI[\img]{l}(L){=}\hbar^2l(l{+}1)\rhoI[\img]{l}(L),
\end{gather}
The general solution of the fourth-order differential equation \eqref{app1.-L^2r=hl(l+1)r} depends on 4 free parameters $c_{\alpha,\beta}$~($\alpha,\beta{=}{\pm}1$):
\begin{gather}
\rhoI[\img]{l}(L){=}\sum_{\alpha,\beta{=}\pm1}c_{\alpha,\beta}\frac{e^{-\frac{4 L}{\hbar }}}{\left({L}/{\hbar}\right)^{1{-}\xi}} \Laguerre_{(2l{+}1)\beta{-}\alpha\xi{-}\frac{1}{2}}^{(2\alpha\xi)}\left[\frac{8L}{\hbar}\right],
\end{gather}
where $\Laguerre_i^{(j)}$ denotes the associated Laguerre polynomial. The particular solution of interest satisfies the conditions \eqref{app1.-[r_1,r_2]=0} and $\rhoI[\img]{l}(L)|_{L{\to}\infty}{\to}0$:
\begin{gather}
\rhoI[\img]{l}(L){=}\sqrt{\frac{(2 l{+}1) 2^{6 \xi {+}1}}{\Gamma(2 l{-}\xi {+}\frac{3}{2}) \Gamma (2 l{+}\xi {+}\frac{3}{2})}} \frac{e^{{-}\frac{4 L}{\hbar}}}{\sqrt{\pi }\hbar^{\frac32}} \left(\frac{L}{\hbar}\right)^{\xi {-}1} {\times}\notag\\
{ \left(\vphantom{\int}U({-}2 l{+}\xi {-}{1}/{2},2 \xi {+}1,{8 L}/{\hbar}){-}\right.}\notag\\
{\left.\frac{\Gamma(2 l{+}\xi {+}\frac{3}{2})}{\Gamma ({-}2 l{+}\xi {-}\frac{1}{2})} U(2 l{+}\xi {+}{3}/{2},2 \xi {+}1,{8 L}/{\hbar})\right)}, \label{app1.-i-solution}
\end{gather}
where the $U(a,b,z)$ are the confluent hypergeometric functions of the 2-nd kind and $l{>}{-}\frac12$. Eq.~\eqref{app1.-i-solution} allows to find the basis functions $\img\rho_{l,l;l,l}$:
\begin{gather}\label{app1.-|l,l><l,l|}
\img\rho_{l,l;l,l}{=}\frac{1}{(\hbar^{2l}(2 l)!)^2}\left(\imgl L_1{-}i\imgl L_2\right)^{2l}\left(\imgr L_1{+}i\imgr L_2\right)^{2l}\rhoI[\img]{l}.
\end{gather}
In the special case of $\xi{=}1/2$ only one of $c_{\alpha,\beta}$ is nonzero:
\begin{gather}\label{app1.-i-solution;xi=1/2}
\rhoI[\img]{l}(L)|_{\xi{=}\frac12}{=}\frac{4}{\hbar\sqrt{\pi L}}\frac{(-1)^{2 l} e^{{-}\frac{4L}{\hbar }} {\Laguerre}_{2 l}^{(1)}\left(\frac{8L}{\hbar }\right)}{\sqrt{\frac{L}{\hbar }}};~~l{=}0,\frac12,1,\frac32,...,
\end{gather}
Using \eqref{app1.-i-solution;xi=1/2} and the relation:
\begin{gather}
e^{{-}\gamma x}{=}\sum_{i=0}^\infty \frac{\gamma^i}{(1{+}\gamma)^{i+\alpha+1}} {\Laguerre}_i^{(\alpha)}(x),
\end{gather}
one can show that $\rhoI[\img]{}|_{\xi{=}\frac12}{=}\frac{1}{\hbar\sqrt{\pi L}}$.

\section{Classical and quantum description of the rigid body dynamics in terms of quaternions\label{@SEC.-app0}}
For the sake of completeness of the presentation, in this Appendix we review the key formulas of the quaternion algebra and briefly outline the standard representation of the rotational motion in terms of quaternions \eqref{complete.-quaternions_def} (for further details see e.g. \cite{BOOK-Borisov}). For clarity, we use bold symbols $\set x{=}(x_0,x_1,x_2,x_3)$ for quaternion parameters and the symbol $*$ to denote the standard quaternion product:
\begin{gather}\label{app0.-quaternion_multiplication_def}
(\set y*\set x)_k{=}\sum_{i,j{=}0}^3\mq_{k,i,j}y_i x_j,
\end{gather}
where the coefficients $\mq_{k,i,j}$ are defined by Eq.~\eqref{complete.-qtn_multiplication_coeffs}. From the physical point of view, the product \eqref{app0.-quaternion_multiplication_def} represents the result of two successive rotations $\set x$ and $\set y$. For this reason, the product \eqref{app0.-quaternion_multiplication_def} is not commutative. If the norm $||\set x||{=}\sqrt{\sum_{k=0}^3x_i^2}$ of quaternion is not equal to one then each rotation $\set x$ is also accompanied by uniform scaling by factor $||\set x||$. The transformation $\set x^{-1}$ reciprocal to $\set x$ (i.e.~one which restores the initial geometry: $\set x\set x^{-1}{=}\set x^{-1}\set x{=}\set 1$, where $\set 1{=}(1,0,0,0)$) is given by  $\set x^{-1}{=}\frac{\set x^*}{||\set x||^2}$, where the quaternion $\set x^*$ is the conjugate of $\set x$ defined as $\set x^*{=}(x_0,{-}x_1,{-}x_2,{-}x_3)$. The components $\omega_k$ of angular frequency in these notations read as:
\begin{gather}\label{app0.-angular_frequency_components}
\omega_k(\set\lambda,\dot{\set\lambda}){=}2(\set \lambda^{*}*\dot{\set \lambda})_k,
\end{gather}
where $\dot{\set{\lambda}}{=}\der{\set{\lambda}}{t}$. Eq.~\eqref{app0.-angular_frequency_components} allows to determine the generalized momenta $p_{\lambda,k}$ canonically conjugate to $\lambda_k$:
\begin{gather}\label{app0.-canonical_momenta_for_lambda}
\set{p}_{\lambda}{=}\pder{{\rm Lg}}{\dot{\set\lambda}}{=}\set{\lambda}*\tilde{\set{L}},
\end{gather}
where $\rm Lg$ is the classical Lagrangian of the rigid rotor: ${\rm Lg}{=}\frac12\sum_{k=1}^3I_k\omega_k^2(\set\lambda,\der{\set\lambda}{t})$ and $\tilde{\set{L}}{=}(0,L_1,L_2,L_3)$. The canonical expressions for components $L_k$ and $L_k'$ of angular momenta relative to the moving and laboratory frames can be determined by applying to \eqref{app0.-canonical_momenta_for_lambda} a reciprocal transform and  expressions \eqref{complete.-directional_cosines_coeffs} for the direction cosines:
\begin{gather}\label{app0.-L_&_L'-canonical}
L_k{=}\frac12(\set\lambda^**\set p_{\lambda})_k;~~~L_k'{=}{-}\frac12(\set\lambda*\set p^*_{\lambda})_k.
\end{gather}
The quaternions allow one to eliminate the singularities inherent to integration of the dynamical equations in terms of Euler angles. This makes them convenient for a variety of the scientific, engineering, technical and graphics applications \cite{2012-Pujol} including  molecular dynamics simulations \cite{2012-Li,2013-Hidalgo}.

It can be shown \cite{1969-Pierre,1970-Nienhuis} that the passage to the Schrodinger quantum description of rotations can be done in the ordinary way by replacing the $p_{\lambda,k}$ with ${-}i\hbar\pder{}{\lambda_k}$ in \eqref{app0.-L_&_L'-canonical}. Strictly speaking, the variables $\lambda_k$ in this picture represent the angular quaternions up to scaling factors. For this reason, one has to explicitly enforce the correct normalization in the potential part of the Hamiltonian, by replacing the $\lambda_k$ with $\hat\lambda_k{=}\frac{\lambda_k}{||\set\lambda||}$. The associated phase space representation can be trivially obtained using the original Wigner recipe or via the substitutions \eqref{intro.-left_operators_def} and \eqref{intro.-right_operators_def} \cite{1969-Pierre,1970-Nienhuis}.

It is worth  mentioning that although the operators $\hat\lambda_k$ are Hermitian, they can not be associated with quantum-mechanical observable since they include the matrix elements corresponding to fractional changes of angular momentum quantum number $l$ which have never been observed in experiments. The fundamental reason for that is that the corresponding set of operators can be introduced only in an overcomplete configuration space.

\section{Derivation of Eqs.~(\ref{complete.-B:b-ladder_operators}) and (\ref{complete.-B:L_k'})\label{@SEC.-app2}}
It follows from the definition of the quantum-mechanical angular momentum operators that:
\begin{gather}
\left.\exp\left({-}\frac2{\hbar}\Phi \Li[k]'\right)\right|_{\hbar\to0}\img{\rho}{=} {\cal R}_k(\Phi)\img{\rho}\label{app2.-R_k(phi)-def},
\end{gather}
where $\Li[k]'{=}{-}i\frac12(\imgl L_k'{-}\imgr L_k')$ and ${\cal R}_k(\Phi)$ is the classical operator of rotation about axis $\vec e_k'$ by angle $\Phi$, i.e.:
\begin{gather}
{\cal R}_k(\Phi){:}
\begin{cases}
L_s'{\to}\cos((1{-}\delta_{k,s})\Phi)L_s'{+}\sum_{n{=}1}^{3}\epsilon_{s,k,n}\sin(\Phi)L_n';\\
\lambda_s{\to}\sum_{m,n{=}0}^{3}\mq_{s,m,n}r_{k,m}\lambda_n;\\
\qL_s{\to}\sum_{m,n{=}0}^{3}\mq_{s,m,n}r_{k,m}\qL_n,
\end{cases}
\hspace{-8pt}\label{app2.-R_k(phi)-effect}
\end{gather}
where the quaternions $r_{k,m}{=}\delta_{0,m}\cos{\frac{\Phi}{2}}{+}\delta_{k,m}\sin{\frac{\Phi}{2}}$ generate rotations about each of axes $\vec e_k'$ (see Eq.~\eqref{complete.-quaternions_def}). Equations~\eqref{app2.-R_k(phi)-def} and \eqref{app2.-R_k(phi)-effect} allow to determine expressions for the
imaginary parts of $\imgl L_k'$ in different phase spaces $\genphasespace[\Lambda,L']$, $\genphasespace[\Lambda,\qL]$ and $\genphasespace[\xLambda,\qL]$:
\begin{gather}
\hspace{-3pt}\Li[k]'{=}\begin{cases}
\begin{array}{c}
{-}\frac{\hbar}{4}\left(\sum_{m,n{=}0}^{3}\mq_{n,k,m}\Lambda_m\pder{}{\Lambda_n}{+}\right.\\
\left.2\sum_{m,n{=}1}^{3}\mq_{n,k,m}L_m'\pder{}{L_n'}\right),
\end{array}& \genphasespace[\Lambda,L'];\\
{}\frac{\hbar}{4}(\imglr \mu_k{-}\sum_{m,n{=}0}^{3}\mq_{n,k,m}\Lambda_m\pder{}{\Lambda_n}),
& \genphasespace[\Lambda,\qL];\\
{}\frac{\hbar}{4}\imglr \mu_k,
& \genphasespace[\xLambda,\qL],
\end{cases}
\label{app2.-Li_k'}\end{gather}
where $\imglr \mu_k{=}{-}\sum_{m,n{=}0}^{3}\mq_{n,k,m}\qL_m\pder{}{\qL_n}$. Here the transformation \eqref{complete.-(L,Lambda)->(sampi,qL)-transform} was applied to obtain the last line in \eqref{app2.-Li_k'}. The latter relation together with Eqs.~\eqref{complete.-B:L'_Schwinger_oscillator}, \eqref{euler.-right_operators_def} and  commutation relations identical to \eqref{complete.-B:a-ladder_commutators} specify the possible forms of the ladder operators $\imgl b_{\pm}$. The two simplest solutions are given by the operators \eqref{complete.-B:b-ladder_operators} and
\begin{gather}
\imgl b_{\pm}{=}i\qL_{2{\pm}1}{-}\qL_{1{\mp}1};
~~~
\imgl b_{\pm}^{\dagger}{=}\frac12(\pder{}{\qL_{1{\mp}1}}{+}i\pder{}{\qL_{2{\pm}1}}).
\end{gather}
(the latter choice leads to the transpose of \eqref{complete.-B:L_k'} with unbounded right eigenstates and hence should be rejected). The specific choice of constant prefactors in \eqref{complete.-B:b-ladder_operators} is made with the goal to simplify the expressions for quaternion operators (see Appendix~\ref{@SEC.-app4}). Applying the transformation \eqref{complete.-(L,Lambda)->(sampi,qL)-transform} to \eqref{complete.-B:L'_Schwinger_oscillator} and \eqref{complete.-B:b-ladder_operators}, one obtains the following formula for the angular momentum operators in the space $\genphasespace[\Lambda,\qL]$:
\begin{gather}
\imgl L_k'{=}\frac{L_k'(\set\qL)}{L'(\set\qL)}
\left(\left\{\Lqnr[]'(\set{\qL}){-}\Lqnr[](\set{\Lambda})\right\}{+}\Lqnr[](\set{\Lambda})\right){+}\notag\\
\frac{\hbar}{4}\sum_{r{=}0}^3\sum_{s{=}1}^3\mq_{k,r,s}((1{-}\delta_{r,0})\frac{L_r'(\set{\qL})}{L'(\set\qL)}{+}i\delta_{r,0}){\times}\notag\\
(\imglr \mu_s{-}\sum_{m,n{=}0}^{3}\mq_{n,s,m}\Lambda_m\pder{}{\Lambda_n}),\label{app2.-L_k'(lambda-q-basis)}
\end{gather}
where $L_n'(\set\qL)$ is defined by the first of Eqs.~\eqref{complete.-(L,Lambda)->(sampi,qL)-transform} and the operators $\Lqnr[](\set\Lambda){=}\frac{\hbar}8(\sum_{n{=}0}^3(\Lambda_{n}^2{-}\pder{^2}{\Lambda_n^2}){-}4)$ and $\Lqnr[]'(\set\qL){=}\frac{\hbar}4\sum_{n{=}0}^3\qL_n\pder{}{\qL_n}$ are the real parts of the operators $\imgl \Lqn'(\set\qL)$ and $\imgl \Lqn(\set\Lambda)$, such that:
\begin{gather}
\imgl \Lqn'(\set\qL)(\imgl \Lqn'(\set\qL){+}\hbar){=}\imgl L'^2;~~~
\imgl{\Lqn}(\set\Lambda)(\imgl{\Lqn}(\set\Lambda){+}\hbar){=}\imgl L^2.\label{app2.-L_and_L'(lambda-q-basis)}
\end{gather}
Thus, the operators $\imgl \Lqn'(\set\qL)$ and $\imgl{\Lqn}(\set\Lambda)$ are physically equivalent (i.e.~they must produce the same action when applied to any valid physical state $\img\rho$) and are mathematically distinct only due to redundant dimensionality of the phase spaces $\genphasespace[\xLambda,\qL]$ and $\genphasespace[\Lambda,\qL]$. This fact allows one to omit the term in the curly brackets in \eqref{app2.-L_k'(lambda-q-basis)}. Together with Eqs.~\eqref{complete.-(L,Lambda)->(sampi,qL)-transform} and \eqref{app2.-L_and_L'(lambda-q-basis)} it leads to the following set of correspondence relations between phase spaces $\genphasespace[\Lambda,\qL]$ and $\genphasespace[\Lambda,L]$:
\begin{gather}
L_n'(\set\qL){\LR}L_n';~~\imglr\mu_k{\LR}{-}2\sum_{m,n{=}1}^{3}\mq_{n,k,m}L_m'\pder{}{L_n'};\notag\\
\Lqnr[]'(\set{\qL}){-}\Lqnr[](\set{\Lambda}){\LR}0\label{app3.-correspondence:(Lambda,q)<->(Lambda-L)}
\end{gather}
Their substitution into Eq.~\eqref{app2.-L_k'(lambda-q-basis)} leads to Eq.~\eqref{complete.-B:L_k'}. One can directly check that Eqs.~\eqref{complete.-B:L_k'} and \eqref{complete.-B:L_k} are consistent with the condition $\sum_{k{=}1}^3(\imgl L_k'^2{-}\imgl L_k^2){=}0$.

\section{The explicit expressions for the Wigner images of the ladder and quaternion operators in \texorpdfstring{$\genphasespace[\Lambda,L]$}-phase space\label{@SEC.-app4}}
\noindent The aim of this Appendix is to complete the construction of the quantum algebra of the phase space $\genphasespace[\Lambda,L']$ by finding the images of the quaternion operators $\hat\lambda_k$. For convenience of readers who are not interested in the technical details of the derivation we start by providing the final result.

Consider the $\genphasespace[\Lambda,L']$-Wigner representation with the normalization \eqref{app4.-r_l} in \eqref{complete.-B:rho-6D}, the scalar product \eqref{complete.-B:(F,P)} and the images of components of angular momentum defined by Eqs.~\eqref{complete.-B:L_k} and \eqref{complete.-B:L_k'}. The corresponding images of the quaternion operators $\hat\lambda_k$ are given by eqs.~\eqref{app4.-quaternion's_images}, where the compound operators $\imgl{{a}_{\xi}b_{\chi}}$ and $\imgl{{a}_{\xi}^{\dagger}b_{\chi}^{\dagger}}$ ($\xi,\chi{=}{\pm}1$) are specified by Eqs.~\eqref{app4.-compound-lowering}, \eqref{app4.-compound-raising}, \eqref{app4.-S_0} and \eqref{app4.-delta_g}.

To prove this result, we note that the images $\imgl\lambda_k$ can be readily defined in terms of the ladder operators \eqref{complete.-B:a-ladder_operators}, \eqref{complete.-B:b-ladder_operators} (see below). However, the ladder operators $\imgl a_{\pm}^{\dagger}$, $\imgl a_{\pm}$ and $\imgl b_{\pm}^{\dagger}$, $\imgl b_{\pm}$ themselves can exist only in the overcomplete phase space $\{\xLambda,\qL\}$ since their separate application leads to unphysical states with mismatched values of the total angular momentum measured in the laboratory and moving frames. Nevertheless, the phase space $\genphasespace[\Lambda,L']$ can host the compound operators ${\compound[]{\xi}{\chi}}$ and ${\compound[\dagger]{\xi}{\chi}}$, which are free of this problem.

The explicit expressions for the compound operators can be found by applying replacements and substitutions \eqref{complete.-(L,Lambda)->(sampi,qL)-transform} and \eqref{app3.-correspondence:(Lambda,q)<->(Lambda-L)} to their counterparts  $e^{i\frac{\pi}{4}}\imgl{a}_{\xi}\imgl b_{\chi}$, $e^{{-}i\frac{\pi}{4}}\imgl{a}_{\xi}^{\dagger}\imgl b_{\chi}^{\dagger}$ in the $\left\{\xLambda,\qL\right\}$-representation (see eqs.~\eqref{complete.-B:a-ladder_operators}, \eqref{complete.-B:b-ladder_operators}). The result is the following new operators $\imgl g^{-}_{\xi,\chi}$ and $\imgl g^{+}_{\xi,\chi}$:

\begin{widetext}
\begin{gather}
\imgl{g}^{+}_{\xi,\chi}{=}
e^{-i\frac{\pi}4}\frac{1}{2\sqrt{2\hbar L'}}
\left\{\vphantom{\sum_1^3}\right.\hspace{-3pt}
{(\delta_{\chi,{-}\xi}{+}\chi\delta_{\chi,\xi}){\left(\frac{\xi{+}\chi}{2}\left(\xi L'{+}L_3'\right)
{+}\right.}}%\notag\\
\left.\frac{\xi{-}\chi}{2}\left(L_1'{-}i\xi L_2'\right)\right)
\left(\Lambda_3{-}\pder{}{\Lambda_3}{-}i\xi\left(\Lambda_0{-}\pder{}{\Lambda_0}\right)\right){+}\notag\\
\left(\frac{\chi{-}\xi}{2}\left(L_3'{-}\xi L'\right){+}\frac{\xi\chi{+}1}{2}\left(L_1'{+}i\xi L_2'\right)\right){\times}%\notag\\
\left(\Lambda_1{-}\pder{}{\Lambda_1}{-}i\xi\left(\Lambda_2{-}\pder{}{\Lambda_2}\right)\right)
\hspace{-3pt}\left.\vphantom{\sum_1^3}\right\};\label{app4.a+b+}
\end{gather}
\begin{gather}
\imgl{g}^{-}_{\xi,\chi}{=}{-}e^{i\frac{\pi}4}\sqrt{\hbar}\frac{(\chi\delta_{\xi,\chi}+\delta_{\xi,-\chi})}{8\sqrt{2}(L')^{3/2}}{\left(\sum_{n{=}{+}1,{-}1}
%
%\text{PPar2}( n)
\frac{(L_3'{-}\xi n L')(\xi\chi{-}n){+}(n\xi\chi{+}1) \left(L_1'{-}i n\xi L_2'\right)}{2}\right.}\times\notag\\
\left({n\left(\Lambda_{2{-}n}\pder{}{\Lambda_{n{+}1}}{-}\Lambda_{ n{+}1}\pder{}{\Lambda_{2{-} n}}\right){-}i\chi}{+}2iL'{\pder{}{L_3'}}\right)
{\left(\Lambda_{ n{+}1}{+}\pder{}{\Lambda_{ n{+}1}}{-}i\xi(\Lambda_{2{-} n}{+}\pder{}{\Lambda_{2{-} n}})\right)}
{+}\notag\\
\sum_{ n=0}^3i^{\xi n}\left(\sum_{k=0}^3\sum_{s=1}^3\sum_{j=0}^3
{\left\{L_{s}'{\left(\frac{\xi{+}\chi(1{-}(3{-}n)n)}{2}\right)}^2
{\left\{1{-}(1{-}(3{-}s)s)(2(\delta_{j,3{-} n}{+}\delta_{k,3{-} n}){-}1)\right\}}\right.}{+}\right.
{
\xi i^{(1{-}s\chi)}
f_{\xi,\chi,n}}\times\notag\\{
(1{-}\delta_{3,s})(\delta_{j, n}{+}\delta_{k, n})\left.\vphantom{\left(\frac12\right)^2}\right\}\mq_{k,s,j}\Lambda_{j}\pder{}{\Lambda_{k}}{+}
}
{2{L'}\xi
f_{\xi,\chi,n}
(\pder{}{L_2'}{+}i\chi\pder{}{L_1'})}
{\left.{
{+}4i{L'}{\left(\frac{\xi{+}\chi(1{-}(3{-}n)n)}{2}\right)}^2
\left.\vphantom{\sum_{j=0}^3}\right)
(\Lambda_n{+}\pder{}{\Lambda_n})}\right)},\label{app4.a-b-}
\end{gather}
\end{widetext}
where
\begin{gather}
f_{\xi,\chi,n}{=}\frac{1}{2}\left\{\left(L'\xi(1{-}(3{-}n)n)+L_3'\right)(1{-}\xi  \chi(1{-}(3{-}n)n)){+}\right.\notag\\
\left.(\xi\chi{+}(1{-}(3{-}n)n))\left(L_1'{+}i L_2'\xi(1{-}(3{-}n)n)\right)\right\}
\end{gather}
(the phase factors $e^{{\pm}i\frac{\pi}{4}}$ are included for consistency with the generally accepted normalization of the rotational eigenstates $\ket{l,m,k}$). The simplest way to study the effect of the operators \eqref{app4.a+b+} and \eqref{app4.a-b-} on the Weyl symbols is to apply them to the isotropic states \eqref{app3.-rhoI} and \eqref{app4.-r_l}:
\begin{gather}
\sum_{\xi,\chi={-}1,1}\imgl g^{+}_{\xi,\chi}\imgr g^{+}_{\xi,\chi}\rhoI[\img]{l}{=}\mu_l(L')
{4(l{+}\frac12)^2}\rhoI[\img]{l{+}\frac12},
\end{gather}
where $\mu_l(L'){=}\frac{1}{2(l{+}\frac12){+}1}\frac{2L'}{\hbar}$ is an additional factor compared to the expected effect of the compound ladder operator. The correct form of the compound operator can be found by applying to \eqref{app4.a+b+} and \eqref{app4.a-b-} the following transformation, which eliminates this factor:
\begin{gather}
\imgl{{a}_{\xi} b_{\chi}}{=}{\cal S}_0\imgl g^{-}_{\xi,\chi}{\cal S}_0^{{-}1}{=}
(\sqrt{\frac{2L'}{\hbar}}\imgl g^{-}_{\xi,\chi}{+}\imgl{\delta g}^{-}_{\xi,\chi}
)\sqrt{\frac{1}{2\imgl\Lqn/\hbar{+}1}}\label{app4.-compound-lowering},\\
\imgl{{a}_{\xi}^{\dagger}b_{\chi}^{\dagger}}{=}{\cal S}_0\imgl{g}^{+}_{\xi,\chi}{\cal S}_0^{{-}1}{=}
\sqrt{\frac{2\imgl\Lqn}{\hbar}{+}1}\sqrt{\frac{\hbar}{2L'}}\imgl{g}^{+}_{\xi,\chi}\label{app4.-compound-raising},
\end{gather}
where ${\cal S}_0$ is the invariance operator (see Eq.~\eqref{complete.-B:invariance_relation}):
\begin{gather}\label{app4.-S_0}
{\cal S}_0{=}\sqrt{\Gamma\left(\frac{2\imgl\Lqn}{\hbar}{+}2\right)\Gamma\left(\frac{2\imgr\Lqn}{\hbar}{+}2\right)}\left(\frac{2L'}{\hbar}\right)^{{-}\frac{1}{\hbar}(\imgl\Lqn{+}\imgr\Lqn)},
\end{gather}
(here $\Gamma(z)$ is the Euler gamma function) and
\begin{gather}
\imgl{\delta g}^{-}_{\xi,\chi}{=}{\sum _{n=0}^1 \frac{(L'{+}\chi L_3) \delta_{\xi,{-}({-}1)^{n}\chi}{+}(L_1{-}i\chi L_2) \delta_{\xi,({-}1)^{n}\chi}}{i^{{-}n}e^{\frac{{-}i \pi }{4}}4L'}}{\times}\notag\\
(\Lambda _{1{-}n}{+}\pder{}{\Lambda_{1{-}n}}{+}i\xi({-}1)^{n}(\Lambda _{n{+}2}{+}\pder{}{\Lambda _{n{+}2}}))\frac{\imgl\Lqn{+}\imgr\Lqn}{\hbar}.\label{app4.-delta_g}
\end{gather}

To deduce the Wigner images $\imgl \lambda_k$ of the quaternion operators we apply the known relations:
\begin{gather}
\hat\lambda_n{=}
\frac{(-i)^n}{2}D_{\frac{1}{2},\frac{1-(3-n) n}{2}}^{\frac{1}{2}}{+}\frac{e^{i\frac{\pi}{4}((-1)^n{-}1)}}{2}D_{{-}\frac{1}{2},\frac{(3{-}n)n{-}1}{2} }^{\frac{1}{2}};\label{app4.-quaternions_in_WignerDs}
\end{gather}

\begin{gather}
D^{\frac12}_{\delta m,\delta k}\ket{l,m,k}{=}
\sum _{j=l{-}\frac12}^{l{+}\frac12} i^{l{-}j} \sqrt{(2 l{+}1) (2 j{+}1)}{\times}\notag\\
({-}1)^{{\delta m}{+}m{-}({\delta k}{+}k)}
\left(\!\!
\begin{array}{ccc}
 j & \frac12 & l \\
 {-}({\delta k}{+}k) & {\delta k} & k \\
\end{array}
\!\!\right){\times}\notag\\
\left(\!\!
\begin{array}{ccc}
 j & \frac12 & l \\
 {-}({\delta m}{+}m) & {\delta m} & m \\
\end{array}
\!\!\right)
\ket{j,m{+}\delta m,k{+}\delta k},\label{app4.-WignerD-defs}
\end{gather}
where $D_{m,k}^l$ are Wigner D-functions. Since
\begin{gather}
\ket{l{+}\frac12,m{+}\frac{\mu}2,k{+}\frac{\kappa}2}{=}
\frac{\hat a_{\kappa}^{\dagger} \hat b_{\mu}^{\dagger}}{\sqrt{l{+}\kappa k{+}1} \sqrt{l{+}\mu m+1}}\ket{l,m,k},\notag\\
\ket{l{-}\frac12,m{+}\frac{\mu}2,k{+}\frac{\kappa}2}{=}\frac{\hat a_{{-}\kappa} \hat b_{{-}\mu}}{\sqrt{l{-}\kappa k} \sqrt{l{-}\mu m}}\ket{l,m,k},
\end{gather}
($\kappa,\mu{=}{\pm}1$) the action of the operators \eqref{app4.-quaternions_in_WignerDs} can be represented as a bilinear combination of the ladder operators: $\hat\lambda\ket{l,m,k}{=}\sum_{\mu,\kappa}(\hat a_{\kappa}^{\dagger} \hat b_{\mu}^{\dagger}c_1(l,m,k){+}\hat a_{\kappa}\hat b_{\mu}c_2(l,m,k))\ket{l,m,k}$. Since the operators $\hat L^2$, $\hat L_3$ and $\hat L_3'$ commute, the coefficients $c_n(l,m,k)$ may be replaced by the operators $\hat c_n{=}c_n(\frac{\hat L}{\hbar},\frac{\hat L_3'}{\hbar},\frac{\hat L_3}{\hbar})$. After converting the resulting operator into the Wigner representation using Eqs.~\eqref{app4.-compound-lowering}, \eqref{app4.-compound-raising} and \eqref{complete.-B:L}, one finally gets:
\begin{gather}
\imgl\lambda_n{=}
\frac{1}{2\sqrt{\frac{\imgl\Lqn}{\hbar }+\frac{1}{2}}}
\left(e^{\frac{i \pi }{4}} ({-}i)^n {\compound[\dagger]{{-}1}{(3{-}n)n{-}1}}{+}\right.\notag\\
e^{\frac{i \pi}{4}({-}1)^n} {\compound[\dagger]{1}{1{-}(3{-}n)n}}{-}
e^{{-}\frac{i\pi}{4}({-}1)^n} {\compound[]{1}{1{-}(3{-}n)n}}{-}\notag\\
\left.e^{{-}\frac{i \pi }{4}} i^n {\compound[]{{-}1}{(3{-}n)n{-}1}}\right)
\frac{1}{2\sqrt{\frac{\imgl\Lqn}{\hbar }+\frac{1}{2}}}\label{app4.-quaternion's_images}.
\end{gather}

One can readily check that the requested consistency of classical limits 	(\={C}:\ref{complete.-postulate_rev_cl_limit}) holds: $\lim_{\hbar{\to}0}\imgl\lambda_m{=}\lambda_m$. Thus, the Wigner $\{\set \Lambda,\set L\}$-representation defined by Eqs.~\eqref{complete.-B:L_k}, \eqref{complete.-B:L_k'}, \eqref{complete.-B:(F,P)}, \eqref{app4.-compound-lowering}, \eqref{app4.-compound-raising} and \eqref{app4.-quaternion's_images} is self-consistent and complete.

\section{Nasyrov-type phase representation of the rotational motion\label{@SEC.-app5}}
The goal of this Appendix is to derive a Wigner representation in which the quantum generator of motion \eqref{complete.-B:symmetric_top_Liouvillian} for the symmetric top coincides with the corresponding classical Liouvillian $\cal L$. We will depart from the $\{\set\xLambda,\set\qL\}$-representation and convert it into the desired form via a series of transformations. The procedure (and the final expressions for $\imgl \Lqn$ and $\imgl L_k$) in the case of variables $\{\set\Lambda,\set L'\}$ remains the same. However, in this case the expressions of the Bopp operators $\imgl L_k'$ and $\imgl \lambda_m$ are rather cumbersome due to the complicated form of relations $\Lambda_m(\set\xLambda,\set L')$ and will not be presented here.

We start with the fractional Laplace transform of the variables $\xLambda_m$:% $\img \rho(\set\xLambda,\set \qL){\to}\img \rho(\set r,\set \qL)$:
\begin{gather}
\img \rho(\set r,\set L){=}\iiiint_{{-}\infty}^{\infty}\left(\prod_{m{=}0}^3d\xLambda_m\right){\times}\notag\\ e^{\sum_{m=0}^3(\sqrt2\xLambda_m r_m{-}\frac{\xLambda_m^2}2{-}\frac{r_m^2}2)}\img \rho(\set\xLambda,\set L)
\end{gather}
The operators in the spaces $\genphasespace[r,\qL]$ and $\genphasespace[\xLambda,\qL]$ are connected via the correspondence:
\begin{gather}
\xLambda_m\to\frac{r_m{+}\pder{}{r_m}}{\sqrt2};~~~\pder{}{\xLambda_m}\to\frac{\pder{}{r_m}{-}r_m}{\sqrt2}.
\end{gather}
Thus, the effect of $r_m$ and $\der{}{r_m}$ on $\xLambda_m$ is identical to the effect of the ladder operators on the canonic coordinate.
Treating the new variables $r_m$ as proportional to the components of a quaternion, we can formally express them in terms of the associated Euler angles $\alpha$ and $\gamma$:
\begin{gather}
r_0{=}R_1 \cos\left(\frac{\alpha {+}\gamma }{2}\right);~r_1{=}R_2 \cos \left(\frac{\alpha {-}\gamma }{2}\right);\notag\\
r_2{=}R_2 \sin \left(\frac{\alpha {-}\gamma }{2}\right);~r_3{=}R_1 \sin \left(\frac{\alpha {+}\gamma }{2}\right).
\end{gather}
The quantum Liouvillian $\imgl{\cal L}$
 for the symmetric top in the variables $\{R_1,R_2,\alpha,\beta,L_1,L_1,L_3\}$ takes the form \eqref{complete.-B:symmetric_top_Liouvillian} with
\begin{gather}
\imglr\Jqn{=}\frac{\imgl \Lqn{+}\imgr \Lqn{+}\hbar}{2}{=}\frac{1}{4}\hbar\left(R_1 \pder{}{R_1}{+}R_2 \pder{}{R_2}{+}2\right);\notag\\
\imglr\Kqn{=}\frac{\imgl L_3{+}\imgr L_3}2{=}\frac{1}{4} \hbar   \left(R_1 \pder{}{R_1}-R_2 \pder{}{R_2}\right).
\end{gather}
These expressions allow us to further trace the analogy of $R_i$ and $\pder{}{R_i}$ with the ladder operators and consider $\imglr\Jqn$ and $\imglr\Kqn$ as $\pm$ combinations of two harmonic oscillator Hamiltonians. Guided by this analogy, we  make the operator substitution:
\begin{gather}
\pder{}{R_j}{\to}e^{\pder{}{s_j}}\sqrt{s_j};~~~R_j{\to}\sqrt{s_j}e^{{-}\pder{}{s_j}}~~(j{=}1,2),
\end{gather}
which preserves the commutation relation $[\pder{}{R_i},R_j]{=}\delta_{i,j}$. Finally, we introduce the variables:
\begin{gather}
J{=}\frac{\hbar}{4} (s_1{+}s_2{+}2);~~~K{=}\frac{\hbar}{4}(s_1{-}s_2).
\end{gather}
The parameters $\{J,K,\alpha,\beta,\set\qL\}$ constitute the required set of variables in which the Bopp operators of the quantum Liouvillian $\imgl{\cal L}$, the components $\imgl L_k$, the ladder operators \eqref{complete.-B:a-ladder_operators} and the operator $\imgl \Lqn$ defined by \eqref{complete.-B:L(Schrodinger)} take the form:
\begin{gather}
\imgl{\cal L}{=}{-}A J\pder{}{\alpha}{+}(A{-}B)K\pder{}{\gamma};\label{app5.-JKabq-Liouvillian}\\
\imgl L_1{\pm}i\imgl L_2{=}e^{{\mp}i\gamma}\sqrt{\frac{J{\mp}K{-}\frac{\hbar}{2}}{J{\pm}K{-}\frac{\hbar}{2}}}
\frac{\hbar(\pder{}{\gamma}{\pm}\pder{}{\alpha}){\pm}2i(J{\pm}K)}{2} e^{{\pm}\frac{\hbar }{2}\pder{}K};\label{app5.-JKabq-L_1,L_2}\\
\imgl L_3{=}K{-}\frac{1}{2} i \hbar  \pder{}{\gamma};\label{app5.-JKabq-L_3}\\
\imgl \Lqn{=}J-\frac{\hbar}{2}{-}i\frac{\hbar}{2}\pder{}{\alpha};\label{app5.-JKabq-L}\\
\imgl a_{\pm}{=}\frac{ e^{\frac{1}{2} i (\alpha{\pm}\gamma)} \left(\pder{}{\gamma }{\pm}\pder{}{\alpha }{-}i\frac{2 (J{\pm}K)}{\hbar }\right)}{e^{i\frac{\pi}{4}({\pm}1{-}1)}\sqrt{\frac{2 (J{\pm}K)}{\hbar }-1}}e^{\frac{\hbar}{4}(\pder{}{J}{\pm}\pder{}{K})};\label{app5.-JKabq-a-}\\
\imgl a^{\dagger}_{\pm}=\frac{e^{{-}\frac{i}{2}(\gamma{\pm}\alpha)}}{e^{{-}\frac{1}{4} i \pi  ({\pm}1+1)}}\sqrt{\frac{2(J{\pm}K)}{\hbar}{-}1}~e^{{-}\frac{\hbar }{4}(\pder{}{J}{\pm}\pder{}{K})}\label{app5.-JKabq-a+},
\end{gather}
whereas the operators $\imgl b_{\pm}$ and $\imgl b_{\pm}^{\dagger}$ are still defined by Eqs.~\eqref{complete.-B:b-ladder_operators}. It is readily verified that Eqs.~\eqref{app5.-JKabq-Liouvillian}-\eqref{app5.-JKabq-L} have the correct classical limits (recall that the ladder operators \eqref{app5.-JKabq-a-}, \eqref{app5.-JKabq-a+} and \eqref{complete.-B:b-ladder_operators} are specified up to the invariance transform \eqref{complete.-B:invariance_relation}).

One can see that the dynamic master equation \eqref{app5.-JKabq-Liouvillian} exactly coincides with its classical analog. Similarly, the expressions for $\imgl \Lqn$ and $\imgl K$ resemble the canonical Bopp operators \eqref{intro.-left_operators_def} and, in particular, obey the relations:
\begin{gather}
(\imgl \Lqn^n,\rho)_{\idx{W}}{=}(\left(J{-}{\hbar}/{2}\right)^n,\rho)_{\idx{W}};~~(\imgl L_3,\rho)_{\idx{W}}{=}(K^n,\rho)_{\idx{W}},
\end{gather}
so that the associated marginal distributions for the Wigner functions represent the probability distributions for quantities $\Lqn$ and $K$ (cf. Eq.~\eqref{intro.-<q^n>,<p^n>} and subsequent discussion).

The action of operators $e^{{\pm}\frac{\hbar }{4}\pder{}{K}}$ and $e^{{\pm}\frac{\hbar }{4}\pder{}{J}}$  contained in Eqs.~\eqref{app5.-JKabq-L_1,L_2}, \eqref{app5.-JKabq-a-} and \eqref{app5.-JKabq-a+} on any function of variables $J$ and $K$ consist of discrete replacements:
$J{\to}J{\pm}\frac{\hbar}4$, $K{\to}K{\pm}\frac{\hbar}4$. For this reason, the parameters $J$ and $K$ take a discrete set of values, so that the Wigner images $\img \rho_{1,2}$ of the basis functions $\hat\rho_{1,2}{=}\ket{l_1,m_1,k_1}\bra{l_2,m_3,k_2}$ read:
\begin{gather}
\img\rho_{1,2}{\propto}\delta_{\frac{2J}{\hbar},l_1{+}l_2{+}1}\delta_{\frac{2K}{\hbar},k_1{+}k_2}
\end{gather}
(for more details about the explicit form and properties of such a semidiscrete Wigner functions see \cite{1999-Nasyrov}).

\end{document}